\documentclass[aps,prb,twocolumn,citeautoscript,reprint,longbibliography,superscriptaddress]{revtex4-2}
\usepackage{graphicx}
\usepackage{bm}
\usepackage{natbib}
\usepackage{amsmath}
\usepackage{xfrac}
\usepackage{nicefrac}
\usepackage[colorlinks=true, urlcolor=blue, linkcolor=blue, citecolor=blue, pdfborder={0 0 0}]{hyperref}
\usepackage[caption=false]{subfig}
\usepackage{cleveref}
\usepackage{dcolumn}
\usepackage{ulem}

\usepackage{cleveref}
\crefname{figure}{Fig.}{Figs}
\crefname{table}{Table}{Tables}
\crefname{section}{Sec.}{Sections}

\begin{document}

\title{Calculating the spin memory loss at Cu$|$metal interfaces from first principles}

\author{Ruixi Liu}
\affiliation{Center for Advanced Quantum Studies and Department of Physics, Beijing Normal University, 100875 Beijing, China}
\author{Kriti Gupta}
\affiliation{Faculty of Science and Technology and MESA$^+$ Institute for Nanotechnology, University of Twente, P.O. Box 217,
	7500 AE Enschede, The Netherlands}
\author{Zhe Yuan}
\email[Email: ]{zyuan@bnu.edu.cn}
\affiliation{Center for Advanced Quantum Studies and Department of Physics, Beijing Normal University, 100875 Beijing, China}
\author{Paul J. Kelly\thanks{corresponding author}}
\email[Email: ]{P.J.Kelly@utwente.nl}
\affiliation{Center for Advanced Quantum Studies and Department of Physics, Beijing Normal University, 100875 Beijing, China}
\affiliation{Faculty of Science and Technology and MESA$^+$ Institute for Nanotechnology, University of Twente, P.O. Box 217,
	7500 AE Enschede, The Netherlands}

\date{\today}
	
\begin{abstract}
 The role played by interfaces in metallic multilayers is not only to change the momenta of incident electrons; their symmetry lowering also results in an enhancement of the effects of spin-orbit coupling, in particular the flipping of the spins of conduction electrons. This leads to a significant reduction of a spin current through a metallic interface that is quantitatively characterized by a dimensionless parameter $\delta$ called the spin memory loss (SML) parameter, the interface counterpart of the spin-flip diffusion length for bulk metals. In this paper we use first-principles scattering calculations that include temperature-induced lattice and spin disorder to systematically study three parameters that govern spin transport through metallic interfaces of Cu with Pt, Pd, Py (permalloy) and Co:  the interface resistance, spin polarization and the SML. The value of $\delta$ for a Cu$|$Pt interface is found to be comparable to what we recently reported for a Au$|$Pt interface [Gupta {\it et al.}, Phys. Rev. Lett. 124, 087702 (2020)]. For Cu$|$Py and Cu$|$Co interfaces, $\delta$ decreases monotonically with increasing temperature to become negligibly small at room temperature. The calculated results are in good agreement with currently available experimental values in the literature. Inserting a Cu layer between Pt and the Py or Co layers slightly increases the total spin current dissipation at these ``compound'' interfaces. 
\end{abstract}
	
\maketitle
	
\section{Introduction}

Since the discovery of giant magnetoresistance (GMR) in magnetic multilayers \cite{Baibich:prl88, Binasch:prb89, Parkin:prl93}, interfaces have been recognized to play an essential role in the observation of many spintronics phenomena including 
spin-transfer torque \cite{Slonczewski:jmmm96, Berger:prb96, Tsoi:prl98, Waintal:prb00, Xia:prb02, Stiles:prb02, *Ralph:jmmm08, Brataas:prp06, Shao:ieeem21}, 
the spin Hall effect (SHE) \cite{Dyakonov:pla71, Hirsch:prl99, Zhang:prl00, Ando:prl08, Liu:arXiv11, Hoffmann:ieeem13, Sinova:rmp15}, 
spin pumping \cite{Tserkovnyak:prl02a, *Tserkovnyak:prb02b, *Tserkovnyak:rmp05, Saitoh:apl06, Ando:prb08, Mosendz:prl10, Azevedo:prb11, Rojas-Sanchez:prl14, Tao:sca18}, 
the spin Seebeck effect \cite{Uchida:natm10, Xiao:prb10, Guo:prx16}, etc. In particular, the flux of spin angular momentum carried by a spin-polarized current of electrons or by a pure spin current may be significantly reduced at an interface. This loss of spin flux \cite{Fert:prb96b} is described in terms of a dimensionless parameter $\delta$ called the spin memory loss (SML) \cite{Baxter:jap99} and is confirmed in many experimental studies \cite{Kurt:apl02, Eid:prb02, Bass:jpcm07}. When spin transport parameters such as the spin Hall angle and the spin-flip diffusion length (SDL) are being evaluated \cite{Isasa:prb15a, Isasa:prb15b, Nguyen:prl16, Sagasta:prb16, Swindells:prb19, Zhu:apr21}, it becomes critically important to know its numerical value. Neglecting it can lead to severely underestimated values of the SDL $l_{\rm sf}$, which in turn influences the estimated values of the spin Hall angle \cite{Rojas-Sanchez:prl14, LiuY:prl14, Chen:prl15, Wesselink:prb19}. 

In semiclassical transport theory \cite{Valet:prb93}, the transport of a current of spins through an interface is described in terms of an interface resistance $AR_{\rm I}$ and the spin asymmetry $\gamma$, as well as the SML $\delta$. The main way in which these interface parameters are determined is by measuring the GMR in a so-called current-perpendicular-to-the-plane (CPP) configuration. This technique is limited by the small value of the interface resistance between two metals (and its spin dependence) compared to those of typical leads. This problem can be overcome by reducing the sample cross section \cite{Gijs:prl93, *Gijs:ap97}, or by using superconducting leads \cite{Pratt:prl91, *Bass:jmmm99}. At the low temperatures dictated by the superconducting transition temperatures of lead materials like Al or Nb, transport is dominated by disorder, such as interface roughness about which little is usually known for specific samples. Studies based on first-principles scattering theory can model many types of interface disorder \cite{Xia:prb01, Xia:prb02, Xia:prl02, Xu:prl06, Xia:prb06} as well as temperature-induced bulk disorder \cite{LiuY:prb11, *LiuY:prb15} and thereby shed light on the SML and its relationship with microscopic scattering mechanisms \cite{Gupta:prl20}. A number of theoretical studies of the SML have recently been reported for interfaces between nonmagnetic (NM) materials \cite{Belashchenko:prl16, Flores:prb20, Gupta:prl20} and between ferromagnetic (FM) and NM materials \cite{Dolui:prb17, Gupta:prl20, Lim:apl21, Gupta:prb21} with a strong focus on  interfaces between the transition metal Pt and another metal. 

Pt is an important NM metal in spintronics because of its large spin-to-charge conversion efficiency. It is widely used to generate spin currents via the SHE and to detect spin currents via its inverse, the ISHE. 
Pt (and Pd) have very high densities of states at the Fermi energy and are relatively easily magnetized \cite{Gunnarsson:jpf76, Janak:prb77} by proximity to a magnetic material \cite{Huang:prl12, Qu:prl13}. To avoid this happening while incurring minimal attenuation of the spin current, a thin Cu spacer layer is frequently inserted between Pt and magnetic materials \cite{Nakayama:prl13, Caminale:prb16, Emori:apl18} making it of interest to study such ``compound'' interfaces. Because of its long SDL that is estimated to be hundreds of nanometers at room temperature \cite{Bass:jpcm07}, Cu is widely used in nonlocal spin valves as a transport channel for a diffusive spin current \cite{Jedema:nat01, Kimura:prl07, Vila:prl07, Niimi:prl13, Sagasta:prb16, Zhou:prb17, Omori:prb19}. In such studies, the important interfaces are Cu$|$NM and Cu$|$FM interfaces where NM is usually Pt or Pd and FM is Py (permalloy, Ni$_{80}$Fe$_{20}$) or Co. Because of the difficulty of  estimating the SML at interfaces involving Cu it is often simply neglected.

In this paper, we present a systematic study of the transport parameters $AR_{\rm I}$, $\gamma$ and $\delta$ for interfaces comprising Cu and Pd, Pt, Co and Py using first-principles relativistic scattering calculations \cite{Starikov:prl10, *Starikov:prb18} that take into account temperature-induced lattice and spin disorder \cite{LiuY:prb11, *LiuY:prb15} as well as alloy disorder \cite{YuanZ:prl12, *YuanZ:prl14} and lattice mismatch \cite{LiuY:prl14, Gupta:prl20, Gupta:prb21}. The SML parameters for Cu$|$Py and Cu$|$Co interfaces are found to decrease monotonically with increasing temperature and become negligibly small at room temperature. Inserting a thin Cu layer between Pt and Py or Co layers increases the total spin current reduction at the compound Pt$|$Cu$|$FM interface slightly, because of the nonnegligible SML at the Cu$|$Pt interface. 

The rest of this paper is organized as follows. In \cref{sec:method}, we briefly summarize the theoretical methods and provide some technical details of the calculations. The main results are presented and discussed in \cref{sec:results} where we begin by estimating the SDL of Cu, which must be known before we calculate the SML for Cu$|$metal interfaces. This is followed by the results for Cu$|$Pt, Cu$|$Pd, Cu$|$Py and Cu$|$Co interfaces and their dependence on temperature and  interface atomic mixing. A brief summary is given in \cref{sec:conclusion}. In \Cref{sec:app}, the formalism required to determine the SML for NM$|$FM interfaces using a bilayer structure is derived and it is shown to yield the same results for Pt$|$Py interface parameters as were obtained for a Pt$|$Py$|$Pt trilayer in \cite{Gupta:prb21}.

\section{Theoretical methods and computational details}
\label{sec:method}

Most transport experiments in the field of spintronics are interpreted in terms of parameters that are defined in semiclassical transport theory \cite{Valet:prb93}. A typical example is the Valet-Fert model which is used to describe the results of CPP-MR experiments \cite{Stohr:06, Bass:jmmm16} using the bulk transport parameters $\rho$ (resistivity), $\beta$ (conductivity asymmetry), and $l_{\rm sf}$ (SDL) together with the corresponding three interface parameters $AR_{\rm I}$, $\gamma$ and $\delta$. While the calculation of bulk transport parameters using first-principles scattering theory has already been documented \cite{Starikov:prb18, Wesselink:prb19}, determining the interface parameters is less trivial. In this section, we present the formulation \cite{Gupta:prl20} that we use to extract the parameters from the first-principles calculations when combined with a ``layer-averaged local current scheme'' \cite{Wesselink:prb19}. Further details of the formulation are provided elsewhere \cite{Gupta:tbp1, Gupta:prb21}. The results of semiclassical transport theory that we will need are summarized in \Cref{ssec:SDE}. They are applied to an NM$|$NM$'$ interface in \Cref{ssec:nmnmp} and to an FM$|$NM interface in \Cref{ssec:fmnmformulation}. Some technical details of the calculations are given in \Cref{ssec:CD}. 

\subsection{Spin diffusion equation}
\label{ssec:SDE}

In an axially symmetric layered structure, the current flow in the $z$ direction, perpendicular to the interfaces, is described by the spin diffusion equation \cite{Valet:prb93}
\begin{equation}
\frac{\partial^2\mu_s}{\partial z^2}=\frac{\mu_s}{l_{\rm sf}^2},
\label{eq:sde}
\end{equation}
where $l_{\rm sf}$ represents the SDL and $\mu_s$ is the spin accumulation defined as the difference between the spin-up and spin-down chemical potentials, $\mu_s=\mu_\uparrow-\mu_\downarrow$. The spin-dependent chemical potential $\mu_{\uparrow(\downarrow)}$ drives the corresponding current density $j_{\uparrow(\downarrow)}$ according to Ohm's law
\begin{equation}
j_{\uparrow(\downarrow)}(z)=-\frac{1}{e\rho_{\uparrow(\downarrow)}}
\frac{\partial\mu_{\uparrow(\downarrow)}}{\partial z} \label{eq:ohm}
\end{equation}
where $\rho_{\uparrow(\downarrow)}$ is the spin-dependent bulk resistivity and it is assumed that the two spin channels are weakly coupled. 
The general solutions of \eqref{eq:sde} and \eqref{eq:ohm} are 
\begin{equation}
\mu_{s}(z)=Ae^{z/l_{\rm sf}}+Be^{-z/l_{\rm sf}}
\label{eq:solution}
\end{equation}
and
\begin{equation}
j_{\uparrow(\downarrow)}(z)=\frac{1\pm\beta}{2}j\mp\frac{1-\beta^2}{4e\rho l_{\rm sf}}\left(Ae^{z/l_{\rm sf}}-Be^{-z/l_{\rm sf}}\right),
\end{equation}
respectively, where the coefficients $A$ and $B$ are determined by appropriate boundary conditions.
 Here  $\beta=(\rho_\downarrow-\rho_\uparrow)/(\rho_\uparrow+\rho_\downarrow)$ is the bulk spin asymmetry, $j=j_\uparrow+j_\downarrow$ denotes the total current density and $\rho$ is the total resistivity given by $\rho^{-1}=\rho_{\uparrow}^{-1} + \rho_{\downarrow}^{-1}$. We define the normalized spin-current density to be $j_s=(j_\uparrow-j_\downarrow)/(j_\uparrow+j_\downarrow)$ which can be written  
\begin{equation}
\label{eq:js}
j_s(z)=\beta-\frac{1-\beta^2}{2ej\rho l_{\rm sf}}\left(Ae^{z/l_{\rm sf}}-Be^{-z/l_{\rm sf}}\right).
\end{equation}
Equation~\eqref{eq:js} can be used together with the spin current density calculated from first principles to extract $\beta$ and $\l_{\rm sf}$ \cite{Wesselink:prb19} while $\rho$ is determined independently from the scattering matrix using the Landauer formula \cite{Starikov:prl10, Starikov:prb18}. 

\subsection{NM$|$NM$'$ interfaces}
\label{ssec:nmnmp}

For a NM metal or alloy with an equal number of spin-up and spin-down electrons, $\rho_{\uparrow} = \rho_{\downarrow}= 2\rho$, $\beta=0$ and \eqref{eq:js} simplifies to
\begin{equation}
\label{eq:jsnm}
j_s(z)=\frac{1}{2ej\rho l_{\rm sf}}\left(Be^{-z/l_{\rm sf}}-Ae^{z/l_{\rm sf}}\right).
\end{equation}
If a fully polarized spin current is injected from an artificial half-metallic left lead ($z \le 0$) into a diffusive NM$|$NM$'$ bilayer so $j_{s}(0)=1$, then the spin current will decay in both nonmagnetic metals and, provided the bilayer is sufficiently thick, will vanish at the right lead, $j_{s}(\infty)=0$. On each side of the interface, \eqref{eq:jsnm} can be used to fit the calculated $j_s(z)$  to obtain the corresponding SDL $l_i \equiv l^i_{\rm sf}$ for each metal, $i=$NM, NM$'$. Extrapolation of the fitting curves to the interface yields two different values for the spin current at the interface, $j_{s}(z_{\rm I}-\eta) \ne j_s(z_{\rm I}+\eta)$ where $\eta$ is a positive infinitesimal. The corresponding discontinuity, shown in Fig.~\ref{fig3}, represents the interface SML.

In the absence of appropriate boundary conditions for a NM$|$NM$'$ interface, we follow \cite{Eid:prb02, Bass:jpcm07, LiuY:prl14, Rojas-Sanchez:prl14} and model the interface (I) as a fictitious bulklike material with a finite thickness $t$, a resistivity $\rho_{\rm I}$ and SDL $l_{\rm I}$. By doing so, the NM$|$NM$'$ bilayer becomes a NM$|$I$|$NM$'$ trilayer in which the spin current and spin accumulation are everywhere continuous. Taking the limit $t \rightarrow 0$, we recover the conventional interface resistance $AR_{\rm I}$ as well as the SML $\delta$,
\begin{equation}
AR_{\rm I}=\lim\limits_{t \to 0}\rho_{\rm I}t   \,\,\,; \,\,\, \delta=\lim\limits_{t \to 0}t/l_{\rm I}
\end{equation}
where the cross sectional area $A$ should not be confused with the disposable constant in \eqref{eq:jsnm}.
In this way, the discontinuity at the interface is naturally included in the semiclassical spin diffusion theory. The result of imposing continuity on $\mu_s$ and $j_s$ and then taking the limit $t \to 0$ is
\begin{equation}
	\frac{j_{s}(z_{\rm I}-\eta)}{j_{s}(z_{\rm I}+\eta)}=\cosh\delta+\frac{\rho_{\rm NM'}l_{\rm NM'}}{AR_{\rm I}} \, \delta \sinh\delta.
\label{eq:nmnm}
\end{equation}
By calculating the bulk parameters $\rho_{\rm NM'}$ and $l_{\rm NM'}$ and the interface resistance $AR_{\rm I}$ independently, only the single unknown parameter $\delta$ needs to be determined from \eqref{eq:nmnm} which can be straightforwardly solved numerically. The calculation of $AR_{\rm I}$ using the Landauer-B{\"u}ttiker formalism will be presented in \Cref{ssec:CuNM}.

\subsection{FM$|$NM interfaces}
\label{ssec:fmnmformulation}

Unlike NM metals for which $\beta=0$, the asymmetry between the spin-up and spin-down conducting channels in a FM metal leads to $\beta$ saturating to a finite value well inside the ferromagnet, on a length scale of $l_{\rm sf}$. We therefore construct a symmetric NM$|$FM$|$NM trilayer with the origin $z=0$ at the centre of the FM layer and inject an electric current from a NM lead imposing the boundary conditions $j_s(-\infty)=j_s(\infty)=0$. The calculated spin current in the FM metal can be fitted using \eqref{eq:js} to extract $\beta$, $l_{\rm FM}$ \cite{Wesselink:prb19} and $j_{s}(z_{\rm I}-\eta)$ \cite{Gupta:prl20, Gupta:prb21}. In the NM metal, the fitting using \eqref{eq:jsnm} is still applicable resulting in $l_{\rm NM}$ and $j_{s}(z_{\rm I}+\eta)$ where because of the symmetry and without loss of generality, we just consider the rightmost FM$|$NM interface.
Just as we did for the NM$|$NM$'$ interface, we transform the FM$|$NM interface into an FM$|$I$|$NM trilayer with a finite thickness $t$ of the interface region. The only difference is that the FM$|$NM interface resistance is spin dependent and the parameter $\gamma \equiv (AR_{\downarrow}-AR_{\uparrow})/(AR_{\uparrow}+AR_{\downarrow})$ is introduced to characterize the spin polarization. 

Taking the limit $t\to0$ yields the interface discontinuity $j_{s}(z_{\rm I}-\eta) - j_{s}(z_{\rm I}+\eta)\ne 0$ and the analysis \cite{Gupta:prb21} results in two implicit equations 
\begin{widetext}
\begin{subequations}
\label{eq:nmfm}
\begin{eqnarray}
j_{s}(z_{\rm I}-\eta)&=&\gamma-\frac{(1-\gamma^2)\delta}{AR_{\rm I}\sinh\delta}
\left\{\rho_{\rm NM}l_{\rm NM} \, j_{s}(z_{\rm I}+\eta) - \cosh\delta \frac{\rho_{\rm FM}l_{\rm FM}}{1-\beta^2} \tanh\left(\frac{z_{\rm I}}{l_{\rm FM}}\right) \big[\beta-j_{s}(z_{\rm I}-\eta)\big]\right\},\label{eq:nmfm1}\\
j_{s}(z_{\rm I}+\eta)&=&\gamma-\frac{(1-\gamma^2)\delta}{AR_{\rm I}\sinh\delta}
\left\{\rho_{\rm NM}l_{\rm NM} \, j_{s}(z_{\rm I}+\eta)\cosh\delta - \frac{\rho_{\rm FM}l_{\rm FM}}{1-\beta^2}\tanh\left(\frac{z_{\rm I}}{l_{\rm FM}}\right)\big[\beta-j_{s}(z_{\rm I}-\eta)\big]\right\}
\label{eq:nmfm2}
\end{eqnarray}
\end{subequations}
\end{widetext}
which can be solved to yield the remaining unknown variables $\delta$ and $\gamma$ \cite{Gupta:prl20, Gupta:prb21}.

The boundary conditions $j_s(-\infty)=j_s(\infty)=0$ are appropriate for structures with NM materials like Pt, for which $l_{\rm sf}$ is short, embedded between nonmagnetic leads \cite{Gupta:prb21}. When $l_{\rm sf}$ is as long as it is for NM=Cu, it is not possible to treat a thickness of the NM metal so large that the spin current can be assumed to have decayed to zero at the interfaces with the leads. Instead, we consider the injection or detection of a spin-polarized current corresponding to the boundary conditions $j_s(-\infty)$ or $j_s(\infty)$ equal to the current polarization $P$ which can have an arbitrary value in the range $[-1,1]$. For simplicity, a half-metallic ferromagnetic lead is considered in \Cref{sec:app}, where $j_s(-\infty)=\pm1$ is explicitly illustrated. Within the accuracy of the calculations, both sets of boundary conditions lead to the same interface parameters. 

\subsection{Computational details}
\label{ssec:CD}

Our starting point is an electronic structure for the layered metallic system of interest, calculated self-consistently within the framework of density functional theory \cite{Hohenberg:pr64, Kohn:pr65}. The electronic wave functions are expanded in terms of tight-binding linearized muffin-tin orbitals \cite{Andersen:prl84, *Andersen:85, *Andersen:prb86} in the atomic spheres approximation (ASA) \cite{Andersen:prb75}. We use the local density approximation with the exchange-correlation functional parameterized by von Barth and Hedin \cite{vonBarth:jpc72}.  Transport is addressed by solving the quantum mechanical scattering problem for a general two-terminal geometry using Ando's wave-function matching method \cite{Ando:prb91} implemented \cite{Xia:prb06, Zwierzycki:pssb08} with the tight-binding muffin-tin orbital basis. The conductance is calculated directly from the scattering matrix for the scattering region embedded between semi-infinite ballistic leads \cite{Datta:95}. The full quantum-mechanical wave functions are explicitly determined throughout the scattering region from which we calculate position dependent charge and spin currents \cite{WangL:prl16, Wesselink:prb19}. Spin-orbit coupling was neglected in calculating the self-consistent atomic potentials but included in the transport calculations \cite{Starikov:prb18}.

Temperature-induced lattice and spin disorder, alloy disorder and lattice mismatch can be efficiently modelled in ``lateral supercells'' that assume a measure of periodicity in the directions transverse to the transport direction \cite{Starikov:prb18, Wesselink:prb19}. 
The lattice constant of Pt is $a_{\rm Pt}=3.924\,$\AA, that of Pd just 0.8\% smaller, $a_{\rm Pd}=3.891 \,$\AA\ \cite{Ibach:95}. 
Those of Py, Co and Cu are about 10\% smaller but quite similar with $a_{\rm Py}=3.541$~{\AA} \cite{Starikov:prb18}, $a_{\rm Co}= 3.539$~{\AA} \cite{Gupta:19} and $a_{\rm Cu}= 3.615$~{\AA} \cite{Ibach:95}. 
For the Cu$|$Pt, Cu$|$Pd, Py$|$Pt and Co$|$Pt interfaces studied here, an 8$\times$8 interface unit cell of (111) Cu, Py or Co is chosen to match to a 2$\sqrt{13}\times$2$\sqrt{13}$ interface unit cell of Pt or Pd (neglecting the 0.8\% difference) which allows all materials to be kept fcc. The two-dimensional Brillouin zone of the 8$\times$8 supercell of (111) Cu is sampled using $28\times28$ $k$ points and the same $k$ mesh density is used for all the transport calculations. Py is chosen to have its equilibrium lattice constant. To study Cu$|$Py and Cu$|$Co interfaces, Cu must be compressed slightly to match Py, respectively Co; as shown for Au in \cite{Gupta:19, Gupta:tbp1}, this changes the Fermi surface, and hence the transport properties, of the noble metal negligibly. The alloy disorder of bulk Py is modelled by randomly populating lateral supercell sites with Fe and Ni atomic sphere potentials subject to the required stoichiometry \cite{Starikov:prb18}, where the atomic potentials of the alloy are calculated self-consistently within the coherent potential approximation \cite{Soven:pr67, Turek:97}. Interface disorder arising from interface mixing (see Sec.~\ref{ssec:intermix}) is modelled as a number of layers of interface alloy \cite{Xia:prb01, Xia:prb02, Xia:prl02, Xu:prl06, Xia:prb06, Gupta:prl20, Gupta:tbp1}. 

Temperature-induced disorder is modelled in the adiabatic approximation using a ``frozen thermal lattice disorder'' scheme with atoms displaced at random from their equilibrium lattice positions with a Gaussian distribution of displacements. The root mean square displacement $\Delta$ characterizing the distribution is chosen so that the resistivity of the NM metal at a finite temperature is reproduced as detailed in \cite{LiuY:prb11, *LiuY:prb15}. For example, we need a value of $\Delta=0.021a_{\rm Cu}$ to reproduce the room-temperature resistivity of bulk Cu, $1.8\,\mu\Omega\,\mathrm{cm}$. This value of $\Delta$ may be compared to values obtained by populating phonons at 300~K ($0.027a_{\rm Cu}$) \cite{LiuY:prb15}, from room temperature molecular dynamics simulations ($0.023a_{\rm Cu}$) \cite{Yang:prb91} or from 160~K X-ray diffraction measurements ($0.018a_{\rm Cu}$) \cite{Mironets:prb08}. For a FM metal, not only are the atom positions influenced by temperature but also their magnetic ordering. We use a simple Gaussian model of spin disorder parameterized to reproduce the experimental magnetization of the ferromagnet at a given temperature, combined with lattice disorder so that together they reproduce the experimental resistivity as discussed in \cite{LiuY:prb11, *LiuY:prb15, Starikov:prb18}. This spin disorder is equivalent to the Fisher distribution \cite{Fisher:prsa53} at low temperature when the tilting angle measured from the global quantization axis is small \cite{Starikov:prb18}. The numerical convergence has been extensively examined with respect to the maximum angular momentum in the basis, size of the lateral supercell, $k$-mesh sampling of the two-dimensional Brillouin zone, the number of random configurations of lattice and spin disorder, as well as the influence of the three-center integrals in the spin-orbit interaction \cite{Starikov:prb18, Wesselink:prb19, Gupta:prb21, Gupta:tbp1}. Compared to calculations of the conductance, we find that more configurations are needed to reduce the error bars on the local currents to acceptable levels. 

\section{Results and discussion}
\label{sec:results}

Before we calculate the SML for Cu$|$metal interfaces, the SDL of bulk Cu is estimated in \Cref{ssec:lsf}. Cu$|$Pt and Cu$|$Pd interfaces are considered in \Cref{ssec:CuNM}, followed by Cu$|$Py and Cu$|$Co interfaces in \Cref{ssec:CuFM}. The effect of interface atomic mixing for materials with similar atomic volumes is examined in \Cref{ssec:intermix} for Cu$|$Co and Cu$|$Py interfaces. A compound interface obtained by inserting atomic layers of Cu between FM and Pt is studied in \Cref{ssec:Cu_insert}.

\subsection{Estimating $l_{\rm Cu}$}
\label{ssec:lsf}

In the semiclassical theory commonly used to describe spin transport, five bulk parameters $\rho_{\rm NM}$, $l_{\rm NM}$, $\rho_{\rm FM}$, $l_{\rm FM}$, $\beta$ and three interface parameters $AR_{\rm I}$, $\delta$ and $\gamma$ are required to describe transport through the FM$|$NM interface of a bilayer, while the bulk and interface spin polarizations $\beta$ and $\gamma$ vanish for an interface between two NM metals. The numerical techniques required to determine the resistivity of a NM or FM metal, as well as the SDL of most transition metals and alloys using scattering theory are well documented \cite{LiuY:prb11, LiuY:prl14, LiuY:prb15, Starikov:prb18, Wesselink:prb19}. For free-electron like metals like Cu, Ag and Au, $l_{\rm sf}$ may be as large as several hundred nanometers to micrometers and a quantitative estimation requiring the length of the scattering region to be longer than 3-4$\times l_{\rm sf}$ becomes computationally very demanding \cite{Nair:prl21}. 
To estimate $l_{\rm Cu}$ we therefore calculated $\rho_{\rm Cu}$ and $l_{\rm Cu}$ simultaneously at elevated temperatures, as a function of the root-mean-square-displacement $\Delta(T)$, to determine the product $\rho(T) l_{\rm sf}(T)$ that for the Elliott-Yafet mechanism of spin relaxation should be temperature independent and then used the confirmed linear relationship to extrapolate $l_{\rm Cu}$ to the required lower temperature. 
	
\begin{figure}[t]
\includegraphics[width=8.6cm]{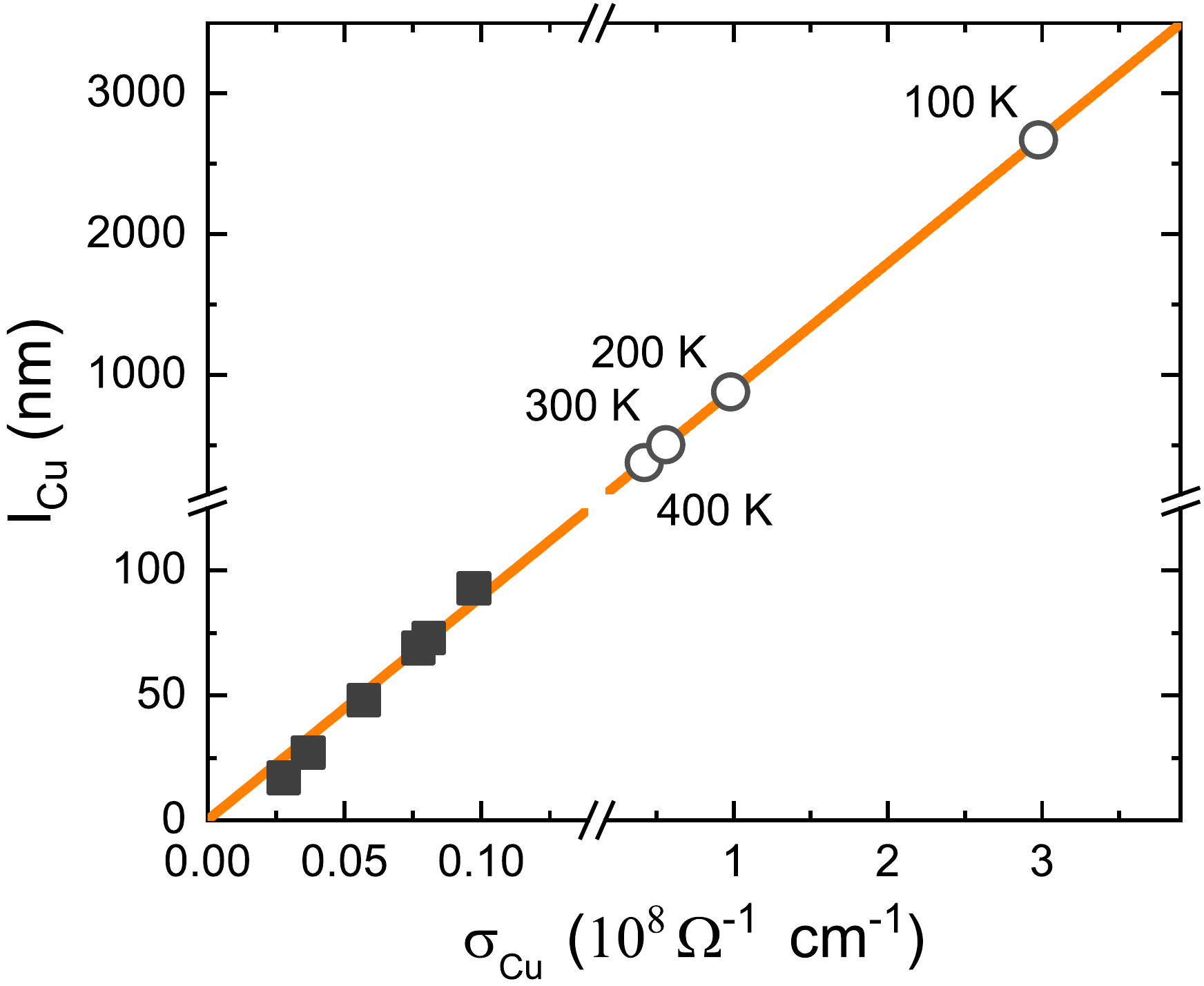}
\caption{Conductivity $\sigma_{\rm Cu}=1/\rho_{\rm Cu}$ and corresponding spin diffusion length $l_{\rm Cu}$ at elevated temperatures (solid squares). The calculations were carried out using 8$\times$8 lateral supercells perpendicular to the spin transport direction. Every point is averaged over 20 random configurations of thermal disorder. From the linear fit to a line that passes through the origin, we find the constant $\rho_{\rm Cu} l_{\rm Cu}=(9.0\pm1.0)\times10^{-15}\ \Omega\ {\rm m}^2$ indicating that the Elliott-Yafet mechanism dominates the spin relaxation. Then the long spin-flip diffusion length at lower temperatures can be determined from the linear relationship using the documented resistivity of Cu at a given temperature (empty circles).}
\label{fig1}
\end{figure}
	
$l_{\rm Cu}(\Delta(T))$ is plotted in \cref{fig1} as a function of the simultaneously determined conductivity for a number of values of $\Delta$ (solid squares); the approximate linearity indicates that the Elliott-Yafet mechanism \cite{Elliott:pr54, Yafet:63} is dominant \cite{Nair:prl21}. The product $\rho_{\rm Cu} l_{\rm Cu}=(9.0\pm1.0) \times 10^{-15}\ \Omega\ {\rm m}^{2}$ is obtained from linear least squares fitting. This linear relationship is extrapolated to the lower temperatures corresponding to the documented conductivity of bulk Cu at 100, 200, 300 and 400 K which are shown as the empty circles in \cref{fig1}. In particular, corresponding to the room temperature resistivity $\rho_{\rm Cu}=1.8\pm0.1\ \mu\Omega$~cm, we estimate a value of $l_{\rm Cu}\sim502$~nm. 

\subsection{Cu$|$Pt and Cu$|$Pd interfaces}
\label{ssec:CuNM}

\begin{figure}[!t]
\includegraphics[width=8.6cm]{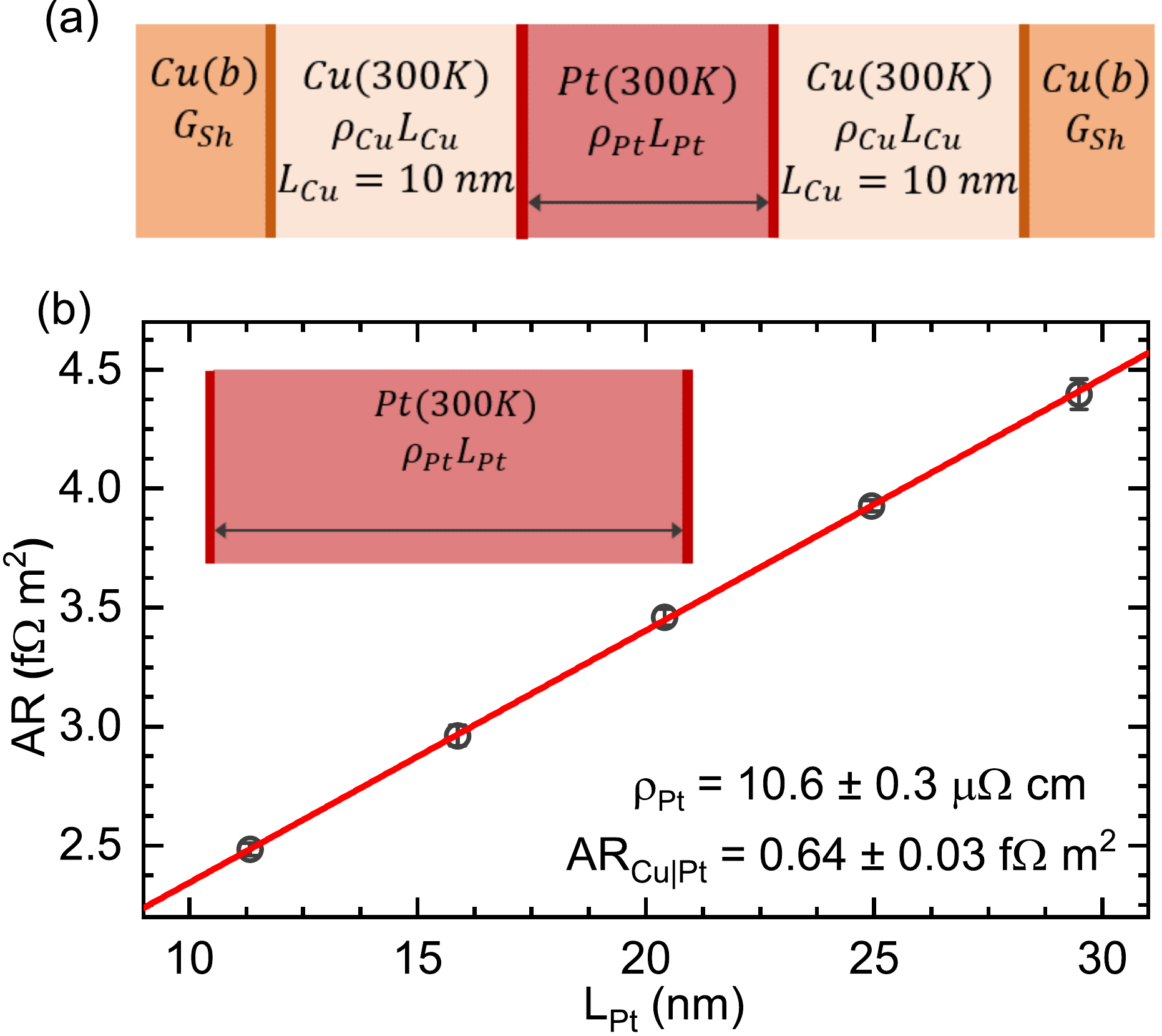}
\caption{(a) Schematic of the Landauer-B{\"u}ttiker scattering geometry for a symmetric, diffusive Cu$|$Pt$|$Cu trilayer sandwiched between ballistic Cu leads. (b) Total resistance \eqref{eq:intres} of this geometry as a function of the Pt thickness $L_{\rm Pt}$ where the Sharvin conductance of the Cu leads, the interface resistance between the ballistic Cu leads and diffusive Cu, and the resistance of a length $2L_{\rm Cu}$ of diffusive Cu have been subtracted leaving just $2AR_{\rm Cu|Pt}+\rho_{\rm Pt}L_{\rm Pt}$. The data points at every thickness are the average of seven configurations of random thermal disorder. See the text for more details.
\label{fig2}}
\end{figure}

We begin by examining the spin memory loss at the Cu$|$Pt interface that commonly appears in nonlocal spin-valve, spin-pumping and spin-Seebeck experiments. We then compare our results when Pt is replaced by the isoelectronic and isostructural but lighter element Pd. 

Before addressing the SML, the Cu$|$Pt interface resistance needs to be determined using the standard Landauer-B{\"u}ttiker formalism \cite{Datta:95, Imry:02}. As sketched in \cref{fig2}(a), we do this by sandwiching a symmetric, diffusive Cu$|$Pt$|$Cu trilayer between ballistic Cu leads in a two-terminal scattering configuration where element specific parameters $\Delta_i$ are used to reproduce the room-temperature resistivities of bulk Cu and Pt, respectively. The total calculated area resistance product for a cross-section $A$ of the scattering geometry can be written as 
\begin{eqnarray}
AR_{\rm total}=1/G_{\rm Sh}&+&2AR_{\rm Cu|Lead}+2\rho_{\rm Cu}L_{\rm Cu} \nonumber \\
                           &+&2AR_{\rm Cu|Pt}+\rho_{\rm Pt}L_{\rm Pt}, 
\label{eq:intres}
\end{eqnarray}
where $G_{\rm Sh}$ denotes the Sharvin conductance of the ballistic Cu leads, $AR_{\rm Cu|Lead}$ is the resistance of the interface between the ballistic Cu lead and diffusive Cu, $2\rho_{\rm Cu}L_{\rm Cu}$ with $L_{\rm Cu}=10\,$nm is the resistance of a length $2L_{\rm Cu}= 20 \,$nm of diffusive Cu \footnote{An explicit numerical check showed that $L_{\rm Cu}=10\,$nm is enough to reproduce the linear dependence of the resistance on the Cu length and therefore a longer $L_{\rm Cu}$ would not alter the final value of the SML that we are interested in.},
 $\rho_{\rm Pt}L_{\rm Pt}$ is the resistance of a length $L_{\rm Pt}$ of diffusive Pt and $AR_{\rm Cu|Pt}$ is the sought-after Cu$|$Pt interface resistance. By repeating the calculations without the central Pt layer, we obtain the contribution from the first three terms on the right hand side of \eqref{eq:intres} separately. This is subtracted from the total resistance in \eqref{eq:intres} leaving us with $2AR_{\rm Cu|Pt} + \rho_{\rm Pt}L_{\rm Pt}$ that is plotted in \cref{fig2}(b) as a function of $L_{\rm Pt}$ where each data point is obtained by averaging over seven configurations of random thermal disorder. The slope of the linear least squares fit reproduces the resistivity of bulk Pt at room temperature while the intercept results in the interface resistance $AR_{\rm Cu|Pt}=0.64\pm0.03\ {\rm f}\Omega\ {\rm m}^2$.

\begin{figure}[t]
\includegraphics[width=8.6cm]{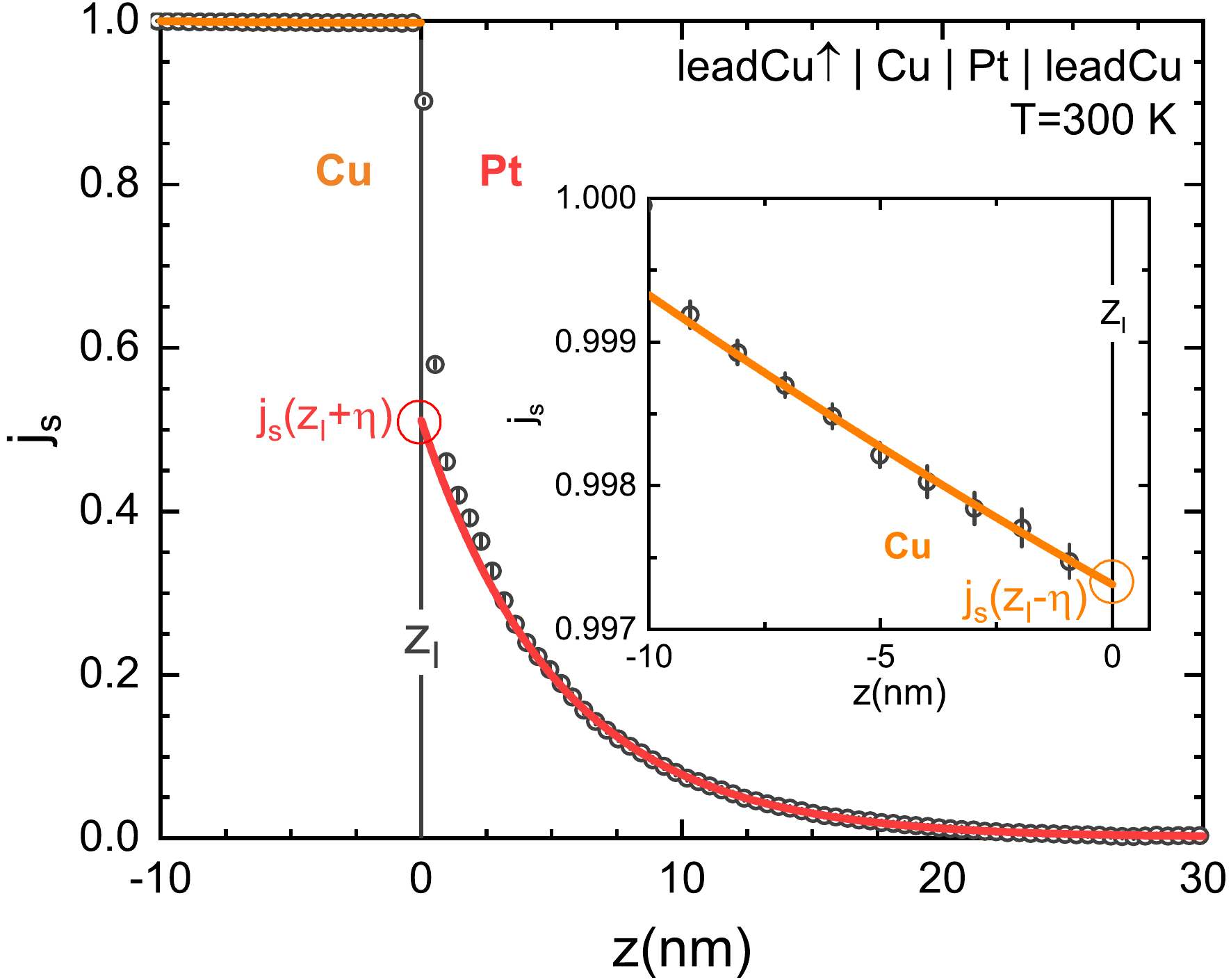}
\caption{Calculated spin current flowing through a diffusive Cu(10 nm)$|$Pt(30 nm) bilayer at 300 K, where a fully polarized spin current is injected from an artificial half-metallic Cu left lead. Averaged spin currents $j_s$ are shown as grey circles. The error bars correspond to the root-mean-square deviation of twenty random configurations of thermal lattice disorder. The solid lines are fits using \eqref{eq:jsnm} in Cu (orange) and Pt (red), respectively. $j_{s}(z_{\rm I}-\eta)$ and $j_{s}(z_{\rm I}+\eta)$ indicate the values of the bulk spin currents extrapolated to the interface, $z_{\rm I}$, from the Cu and Pt sides, respectively. Inset: close-up of the spin current in the Cu layer.}
\label{fig3}
\end{figure}
	
$l_{\rm Pt}$ \cite{Wesselink:prb19, Nair:prl21}, $l_{\rm Cu}$ and $AR_{\rm Cu|Pt}$ have now been calculated, $\rho_{\rm Cu}$ and $\rho_{\rm Pt}$ have their experimental values by construction, leaving just $j_{s}(z_{\rm I}+\eta)$ and $j_{s}(z_{\rm I}-\eta)$ to be determined before $\delta_{\rm Cu|Pt}$ can be evaluated using \eqref{eq:nmnm}. To do so, we inject a fully polarized spin current from an artificial half-metallic Cu lead \cite{Starikov:prb18, Wesselink:prb19} into a diffusive, room temperature (RT: 300 K) Cu(10~nm)$|$Pt(30~nm) bilayer. The $z$-dependent spin current density is plotted in \cref{fig3}, where each data point corresponds to $j_s$ averaged over an atomic layer and over twenty random configurations of thermal lattice disorder. On both sides of the interface, the spin current can be fitted very well using \eqref{eq:jsnm} as shown by the solid orange and red lines in \cref{fig3}. As expected from the large value of $l_{\rm Cu}$, the spin current decays extremely slowly in Cu, as shown in the inset. To reduce the uncertainty from fitting, we use the room temperature value $l_{\rm Cu}\sim 502$~nm yielding very good agreement with the calculated data. The spin current entering Pt exhibits an exponential decay which can be characterized by a value of $l_{\rm Pt}=5.3 \pm 0.1 \,$nm consistent with the value $5.26\pm 0.04$~nm reported previously in \cite{Gupta:prl20}. The fitted semiclassical description of the spin current exhibits a substantial discontinuity at the Cu$|$Pt interface corresponding to the interface SML, \cref{fig3}. Using the values $j_{s}(z_{\rm I}+\eta)=0.997$ and $j_{s}(z_{\rm I}-\eta)=0.512$ obtained by extrapolation, $\rho_{\rm Pt} l_{\rm Pt}=0.6~{\rm f}\Omega~{\rm m}^2$, and $AR_{\rm Cu|Pt}=0.64\pm0.03~{\rm f}\Omega~{\rm m}^2$, we finally estimate the SML to be $\delta_{\rm Cu|Pt}=0.77 \pm 0.04$ for the Cu$|$Pt interface. This value of $\delta$ is very close to and only slightly smaller than the value of 0.81 we reported for the relaxed Au$|$Pt interface and substantially larger than the value 0.62 we found for the compressed Au$|$Pt interface \cite{Gupta:prl20} indicating the importance of taking lattice mismatch into account, \cref{TableI}.

By repeating the complete procedure for the Cu$|$Pd interface and neglecting the 0.8\% smaller lattice constant of Pd, we find $\delta=0.45 \pm 0.03$ at 300 K that is smaller than the SML for the Cu$|$Pt interface because of the weaker spin-orbit interaction in Pd. This value for the Cu$|$Pd interface is also consistent with the SML values of 0.43 and 0.63 calculated for the ``commensurate'' (with Au compressed to match to Pd) and ``incommensurate'' Au$|$Pd interfaces \cite{Gupta:prl20}, respectively. We attribute this agreement to the very similar electronic structures of Cu and Au, where the free-electron-like $s$ band dominates transport at the Fermi level; the filled 3$d$ or 5$d$ bands are well below the Fermi energy so the difference in spin-orbit coupling strength between Cu and Au does not play a significant role.

\begin{table}[t]
\caption{Spin memory loss $\delta$ for Cu$|$Pd, Cu$|$Pt, Au$|$Pd and Au$|$Pt (111) interfaces. For Au$|$Pd and Au$|$Pt, we consider ``commensurate'' interfaces where Au is compressed so its lattice constant matches that of Pd/Pt and ``incommensurate'' where the lattice mismatch between fully relaxed Au and Pd/Pt is accommodated using the large lateral supercells discussed in \Cref{ssec:CD}. }
\begin{ruledtabular}
\begin{tabular}{lcccc}
                   & \multicolumn{2}{c}{Pd}        & \multicolumn{2}{c} {Pt} \\
                     \cline{2-3}                      \cline{4-5} 
$\delta$           & Compressed    & Relaxed       &  Compressed   & Relaxed    \\ 
\hline
Cu                 &    ---        & $0.45\pm0.03$ &  ---          & $0.77\pm0.04$ \\
\hline
Au \cite{Gupta:19} & $0.43\pm0.02$ & $0.63\pm0.02$ & $0.62\pm0.03$ & $0.81\pm0.05$ \\
\end{tabular}
\end{ruledtabular}
\label{TableI}
\end{table}

\begin{table}[b]
\caption{Experimental and calculated interface resistance $AR_{\rm I}$ and spin memory loss $\delta$ for Cu$|$Pt and Cu$|$Pd interfaces.}
\begin{ruledtabular}
\begin{tabular}{lllcrl}
        & $AR_{\rm I}(\rm f\Omega\,{\rm m^2})$     & $~~~~~\delta$ &	T(K)  & ~Method & Ref. \\
\hline
Cu$|$Pt & $0.75\pm0.05$                          & $0.9\pm0.1$	 & ~~~4.2 & CPP-MR  & \cite{Bass:jmmm16}   \\
        & ——                                     & 0.95          & ~~~7   & CPP-MR  & \cite{Freeman:prl18} \\
        & ——                                     & 0.89          &  295   & CPP-MR  & \cite{Freeman:prl18} \\
        & $0.64\pm0.03$                          & $0.77\pm0.04$ &  300   &	\multicolumn{2}{c}{This work}          \\	
\hline
Cu$|$Pd & $0.45\pm0.05$ & $0.24^{+0.06}_{-0.03}$ & ~~~4.2        & CPP-MR &	\cite{Bass:jmmm16}  				  \\
        & 0.76                                   & 0.43          & ~~~0   & Ab initio	& \cite{Belashchenko:prl16} \\
        & $0.50\pm0.03$                          & $0.45\pm0.03$ & 300    &	 \multicolumn{2}{c}{This work}  	\\
\end{tabular}
\end{ruledtabular}
\label{TableII}
\end{table}

We compare our calculated interface resistances and SML parameters with experimental and theoretical values available in the literature in \cref{TableII}. For the Cu$|$Pt interface, our calculated values of $AR_{\rm I}$ and $\delta$ are somewhat smaller than the experimental values. Experimentally, the influence of temperature on the SML of a NM$|$NM$'$ interface is found to be weak \cite{Freeman:prl18} which is consistent with our theoretical findings \cite{LiuY:prl14, Gupta:prl20}. This is because the main scattering mechanism at the interface is the abrupt variation in the atomic potentials experienced by conduction electrons \cite{LiuY:prl14}. For the Cu$|$Pd interface, the value we calculate for $\delta$ is in perfect agreement with that calculated by Belashchenko {\it et al.} \cite{Belashchenko:prl16} but is larger than the only reported, low-temperature experimental value \cite{Bass:jmmm16}. To make more progress, structural properties need to be correlated with transport measurements. We identify a lack of structural characterization of interfaces on the atomic level as a major stumbling block to developing more comprehensive understanding. The theoretical description we have presented can be applied to considerably more complex structural models.

\subsection{Cu$|$FM interfaces}
\label{ssec:CuFM}

We proceed to study the SML at the interfaces between Cu and the FM metals Py and Co. The exchange interaction in FM metals automatically generates a spin polarization of the conduction electrons so that we do not need to artificially inject a spin-polarized current from the half-metallic lead as in the previous section. 

\subsubsection{Cu$|$Py}

\begin{figure}[t]
\includegraphics[width=8.6cm]{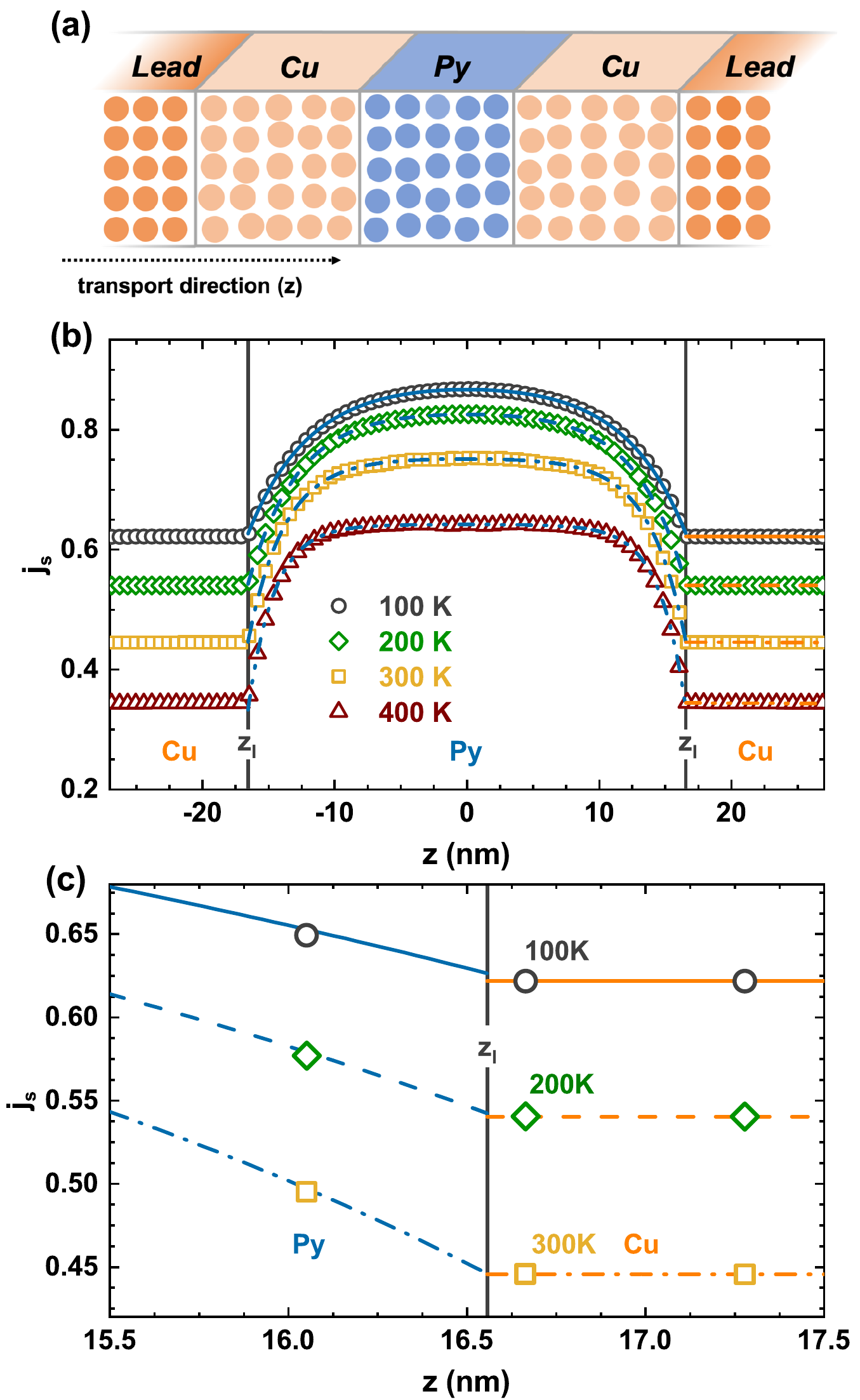}
\caption{(a) Schematic illustration of the scattering configuration for a thermally disordered Cu$|$Py$|$Cu trilayer sandwiched between ideal Cu leads. (b) Calculated spin current $j_s(z)$ in a Cu$|$Py$|$Cu multilayer at 100 K (circles), 200 K (diamonds), 300 K (squares) and 400 K (triangles). The solid, long-dashed, dash-dotted and dotted curves are fits to \eqref{eq:js} in Py (blue) and \eqref{eq:jsnm} in Cu (orange) at 100, 200, 300 and 400 K, respectively. (c) Magnified plot of the right Py$|$Cu interface.}
\label{fig4}
\end{figure}

Instead, we consider a Py thin film that is sandwiched between two diffusive Cu films, as sketched in \cref{fig4}(a). We pass an electric current through this multilayer and plot the calculated spin current as a function of position in \cref{fig4}(b). According to \eqref{eq:js}, it saturates to the value of the conductivity asymmetry $\beta$ sufficiently deep into Py on a length scale of $l_{\rm Py}$. The saturated plateau value decreases with increasing temperature corresponding to the suppression of $\beta$ by magnons \cite{LiuY:prb15}. A lower but nonzero spin polarization is seen in the diffusive Cu layers. The spin current in Py and Cu can be fitted using equations \eqref{eq:js} and \eqref{eq:jsnm}, respectively, as shown by the lines in \cref{fig4}. Unlike the corresponding Pt$|$FM$|$Pt case \cite{Gupta:prl20, Gupta:prb21}, the spin current does not show a significant discontinuity at the Py$|$Cu interface. A magnified plot about the right interface in \cref{fig4}(c) shows that a very small discontinuity appears only at low temperature. In particular, the quantitative calculation results in $\delta=0$ at 300 K. After taking into account all of the uncertainties in the calculated transport parameters entering \eqref{eq:nmfm}, we obtain an estimate of 0.09 as the upper limit of $\delta$ for the Cu$|$Py interface at room temperature. The interface spin asymmetry coefficient is estimated to be $\gamma=0.97\pm0.01$, \Cref{TableIII}. This high value can be understood in terms of the very similar majority-spin potentials of fcc Cu and of Fe and Ni in fcc Py~\cite{Starikov:prb18}, where the majority-spin 3$d$ bands are calculated to be filled.
The Cu$|$Py interface then scatters majority-spin electrons only very weakly. In contrast, the minority-spin potentials (and electronic structures)  change abruptly at the interface resulting in strong scattering. A highly asymmetric interface conductance is the result. 

The very long SDL of Cu introduces two technical difficulties in the above calculations: (i) the value of $l_{\rm Cu}$ we use is extrapolated from high temperatures and (ii) the diffusive Cu layers attached to Py are not thick enough for the boundary condition $j_s(\pm\infty)=0$ to be satisfied as required by the formulation of \cref{ssec:fmnmformulation}. 

We first examine the robustness of our computational framework with respect to the value of $l_{\rm Cu}$. In addition to using the value of $l_{\rm sf} \sim 502\,$nm estimated by extrapolation in \cref{fig1}, we also use literature values of $l_{\rm Cu}$, found in room-temperature experiments and listed in \cref{TableIII}, to calculate $\delta$ and $\gamma$ for the Cu$|$Py interface. Using these values in \eqref{eq:nmfm} always yields a vanishingly small SML at room temperature. This is insensitive to the particular value of $l_{\rm Cu}$ used and only its upper bound shows a slight variation. The value of $\gamma$ we estimate is relatively sensitive to the value of $l_{\rm Cu}$ used; a longer $l_{\rm Cu}$ results in a larger value of $\gamma$. This is because the normalized spin currents $j_{s}(z_{\rm I}+\eta)$ on the Cu side and $j_{s}(z_{\rm I}-\eta)$ on the Py side of the interface have values between 0 and 1 and the large product $\rho_{\rm Cu} l_{\rm Cu}$ in \eqref{eq:nmfm} must be compensated by the factor $1-\gamma^{2}$ to avoid exceeding these bounds.  In particular, we see that in the limit $l_{\rm Cu} \rightarrow \infty$, $\gamma=1$. Therefore, a reasonable value of $l_{\rm Cu}$ is needed to estimate $\gamma$, the interface resistance asymmetry.

\begin{table}[t]
\caption{Comparison of the values of $l_{\rm Cu}$ (nm), $\rho_{\rm Cu} l_{\rm Cu}$ (f$\Omega \,$m$^2$), $\gamma$ and $\delta$ we calculate for the Cu$|$Py (111) interface at room temperature, ``This work'', with values of $\delta_{\rm Cu|Py}$ and $\gamma_{\rm Cu|Py}$ we estimate using values of $l_{\rm Cu}$ and $\rho_{\rm Cu} l_{\rm Cu}$ reported from various experiments. 
LNL/M: Lateral non-local MR with metallic contacts and no other special conditions. 
LNL/+: Lateral non-local MR with an extra strip or strips across the NM-metal. 
LNL/C: Lateral non-local MR using a cross-geometry for the NM-metal. 
CPP-NP: CPP-MR using electron-beam lithography produced nanopillar trilayers. 
CPP-NW: CPP-MR using electrodeposited nanowire multilayers. 
The value of $l_{\rm Cu}$ that we extrapolate from the first-principles calculations at high temperatures shown in Fig.~\ref{fig1} is also listed for comparison. 
}
\begin{ruledtabular}
\begin{tabular}{lllcll}
T(K) & Method	
                                              & $l_{\rm Cu}$(nm)	
                                                            & $\rho_{\rm Cu} l_{\rm Cu}$	
                                                                         & $\delta_{\rm Cu|Py}$
                                                                                       & \multicolumn{1}{c} {$\gamma_{\rm Cu|Py}$}  \\
\hline
300  & CPP-NW \cite{Doudin:jap96}             &	36$\pm$14  & 0.4$-$3     & $0^{+0.07}$ & 0.42$\pm$0.35   \\
300	 & LNL/M \cite{Ji:apl06}                  & $110$      & $4$         & $0^{+0.09}$ & 0.85$\pm$0.03   \\
293  & CPP$-$NP \cite{Albert:prl02}           &	170$\pm$40 &  —          & $0^{+0.08}$ & 0.91$\pm$0.06   \\
293  & LNL/C \cite{Jedema:nat01,Jedema:prb03} &	350$\pm$50 & 10          & $0^{+0.09}$ & 0.96$\pm$0.02   \\	
293  & LNL/M;                                 &            &             &             &                 \\
     & LNL/+ \cite{Kimura:prb05}              & 500        & 11          & $0^{+0.09}$ & 0.97$\pm$0.01   \\
293  & LNL/M \cite{Maekawa:06}                & 700        & 15          & $0^{+0.09}$ & 0.98$\pm$0.01   \\	
\hline
300  & This work                              & 502        & 9.0$\pm$1.0 & $0^{+0.09}$ & 0.97$\pm$0.01   \\
\end{tabular}
\end{ruledtabular}
\label{TableIII}
\end{table}

\begin{figure}[t]
\includegraphics[width=8.6cm]{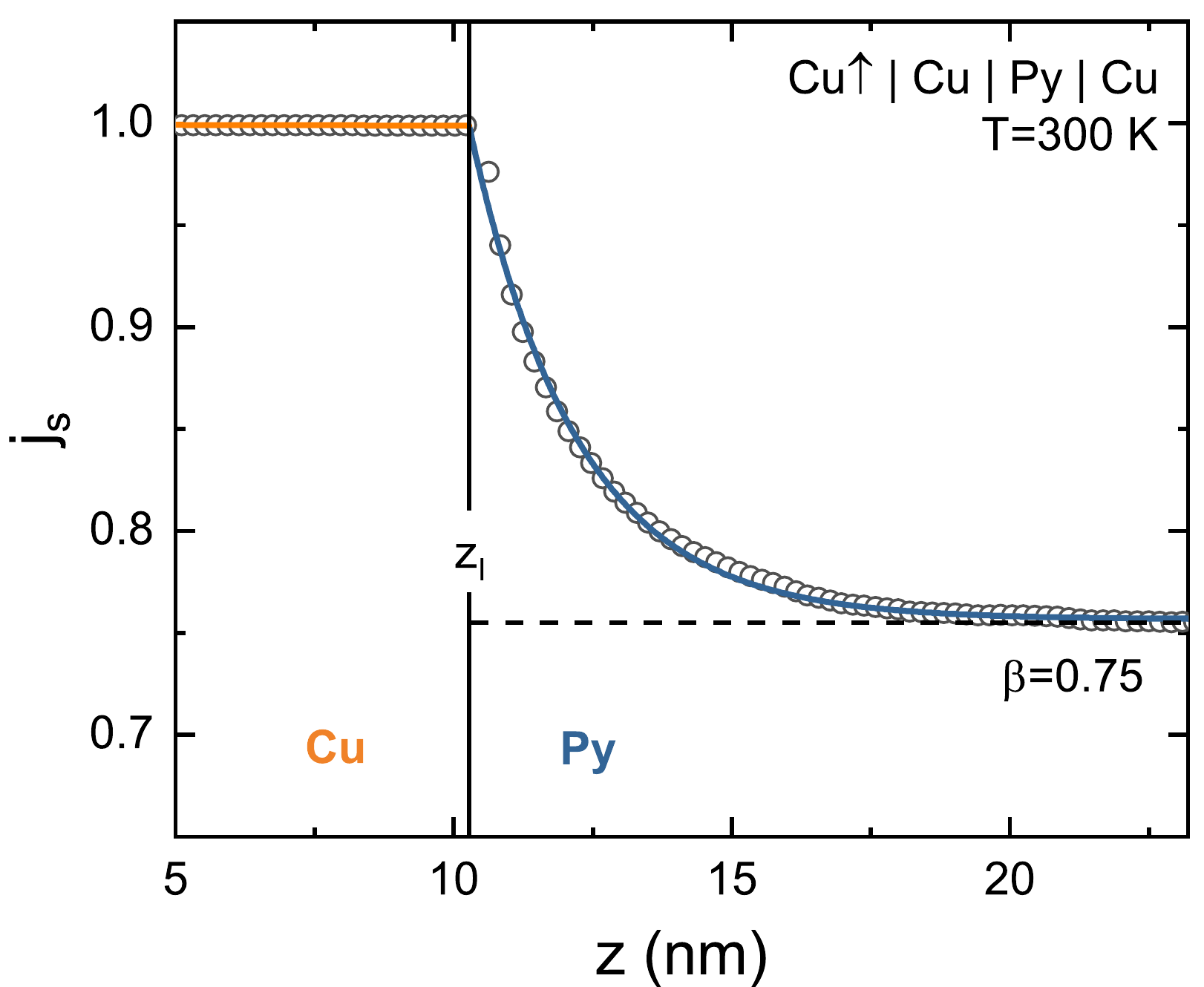}
\caption{Spin current $j_{s}(z)$ calculated for a Cu$|$Py bilayer when a fully spin-polarized current is injected into Cu from the left electrode (empty circles) corresponding to the boundary condition $j_s(-\infty)=1$. The polarization decays in the Cu$|$Py bilayer and approaches the bulk spin polarization well inside Py resulting in $j_s(+\infty)=\beta$. Here the bulk value $\beta=0.75$ is illustrated by the dashed line. The solid lines are fits with equation \eqref{eq:js} to determine $j_s(z_{\rm I}\pm\eta)$ which are used to solve $\delta$ and $\gamma$ in equations \eqref{eq:nmfmv2}.} 
\label{fig5}
\end{figure}

The large value of $l_{\rm Cu} \sim502 \,$nm makes it impossible to construct a scattering region long enough for a Cu$|$FM$|$Cu trilayer to satisfy the boundary condition $j(\pm\infty)=0$ in currently practical calculations. We therefore examine the values of $\delta$ and $\gamma$ we obtain when we apply a different boundary condition that can be strictly complied with in practice. By analogy with the injection of a fully polarized current into the Cu$|$Pt bilayer, we can inject a fully spin-polarized current with the same polarization sign as Py into a Cu$|$Py bilayer; this then satisfies the condition $j_s(-\infty)=1$. When the thickness of Py is much larger than its SDL, the spin current approaches its bulk polarization $\beta$ at $z\rightarrow+\infty$. Following the same procedure we used for the NM$|$NM$'$ interface in \cref{ssec:nmnmp}, we derive the semiclassical diffusion equations for spin transport in a diffusive NM$|$FM bilayer in \Cref{sec:app} to eventually arrive at the two equations 
\begin{widetext}
\begin{subequations}
\small
\label{eq:nmfmv2}
\begin{eqnarray}
j_s(z_{\rm I}-\eta)&=&\gamma-\frac{(1-\gamma^2)\delta}{AR_{\rm I}\sinh\delta}\left\{\frac{\rho_{\rm FM}l_{\rm FM}}{1-\beta^2}\left[j_s(z_{\rm I}+\eta)-\beta\right]-\rho_{\rm NM}l_{\rm NM}\left[{\rm csch}\left(\frac{z_{\rm I}}{l_{\rm NM}}\right)-j_s(z_{\rm I}-\eta)\coth\left(\frac{z_{\rm I}}{l_{\rm NM}}\right)\right]\cosh\delta\right\},\\
j_s(z_{\rm I}+\eta)&=&\gamma-\frac{(1-\gamma^2)\delta}{AR_{\rm I}\sinh\delta}\left\{\frac{\rho_{\rm FM}l_{\rm FM}}{1-\beta^2}\left[j_s(z_{\rm I}+\eta)-\beta\right]\cosh\delta-\rho_{\rm NM}l_{\rm NM}\left[{\rm csch}\left(\frac{z_{\rm I}}{l_{\rm NM}}\right)-j_s(z_{\rm I}-\eta)\coth\left(\frac{z_{\rm I}}{l_{\rm NM}}\right)\right]\right\}.
\end{eqnarray}	
\end{subequations}
\end{widetext}
The spin current $j_s(z)$ calculated in the Cu$|$Py bilayer with the boundary condition $j_s(-\infty)=1$ is plotted in \cref{fig5}. Using \eqref{eq:jsnm} to fit the data calculated in Cu (solid orange line) and \eqref{eq:js} for those in Py (solid blue line) allows us to find the asymptotic values $j_s(z_{\rm I}\pm\eta)=0.998$ at the interface. Since all the bulk parameters $\rho_{\rm NM}$, $l_{\rm NM}$, $\rho_{\rm FM}$, $l_{\rm FM}$, and $\beta$ as well as the interface resistance $AR_{\rm I}$ were already determined independently, equations \eqref{eq:nmfmv2} can be solved resulting in $\delta=0^{+0.05}$ and $\gamma=0.99\pm0.01$. Here the uncertainties in $\delta$ and $\gamma$ are obtained by considering the uncertainties in all the other parameters in \eqref{eq:nmfmv2}. This independent check with an alternative boundary condition confirms the values $\delta=0^{+0.09}$ and $\gamma=0.97\pm0.01$ extracted from Fig.~\ref{fig4} using the trilayer structures. It is worth noting that the thickness of Py in the Cu$|$Py bilayer must be large enough to satisfy the boundary condition $j_s(+\infty)=\beta$. Otherwise the Cu right-hand lead may influence the values of $j_s$ calculated near the Cu$|$Py interface. (Indeed, the influence of the right-hand lead can be seen in the incipient deviation of $j_s(z)$ from $\beta$ for $z>20$nm in \cref{fig5}.)

\begin{figure}[t]
\includegraphics[width=8.6cm]{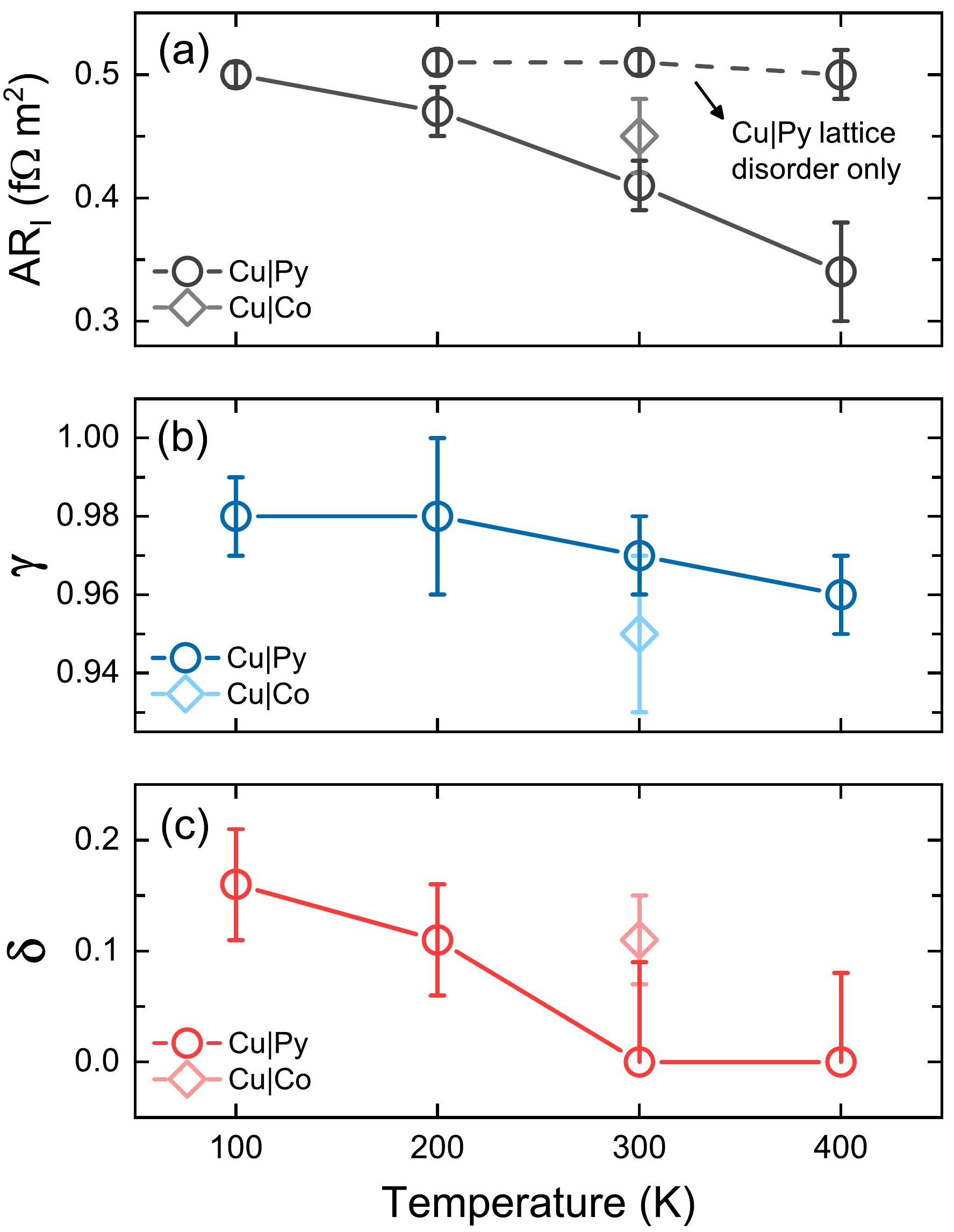}
\caption{Interface parameters $AR_{\rm I}$ (a), $\gamma$ (b) and $\delta$ (c) for Cu$|$Py interface (circles) extracted from the spin currents as a function of temperature. The black circles with a dashed line in (a) show the calculated $AR_{\rm I}$ for the Cu$|$Py interface with only lattice disorder. The corresponding parameters for Cu$|$Co interface at room temperature are also shown (diamonds) for comparison.
\label{fig6}}
\end{figure}

The temperature dependence of $AR_{\rm I}$, $\delta$ and $\gamma$ is shown in \cref{fig6} for the Cu$|$Py interface. All three parameters decrease monotonically with increasing temperature and this decrease can be attributed to the stronger spin disorder of Py at higher temperature by analogy with our findings for the Pt$|$Py interface \cite{Gupta:prl20}. The electron scattering at the Cu$|$Py interface is strongly spin-dependent: the majority-spin channel is highly conductive and, without spin-orbit coupling or spin disorder, the minority-spin channel is much more resistive. Spin disorder allows the spins of conduction electrons to flip at the interface and hence reduces the transmission of majority-spin electrons. At the same time, it creates more transmission channels for minority-spin electrons. Because the minority-spin 3$d$ bands are partially occupied for Py, their state density at the Fermi energy is very large so the increase of the minority-spin conductance is greater than the decrease of that for the majority spins. Therefore the total transmission probability increases with increasing spin disorder; the interface resistance decreases. To confirm this, we repeat the T=200 K, 300 K and 400 K calculations for the Cu$|$Py interface with only lattice disorder and keeping the magnetic moments aligned with the global quantization axis. The values of $AR_{\rm I}$ we calculate are almost independent of temperature as shown in \cref{fig6}. We conclude that spin disorder is the main reason for the temperature-induced decrease of the interface resistance.

\begin{table}[b]
\caption{Interface parameters $AR_{\rm I}$, $\gamma$ and $\delta$ extracted from the currents calculated for Cu$|$Py and Cu$|$Co interfaces from first-principles at various temperatures. 
\label{TableIV}}
\begin{ruledtabular}
\begin{tabular}{llllll}
        & T(K) & $AR_{\rm I}(\rm f\Omega\,{\rm m^2})$ 
                               & ~~~~~$\gamma$ & ~~~~~~$\delta$  \\
\hline
Cu$|$Py & 100  & 0.50$\pm$0.01 & 0.98$\pm$0.01 & 0.16$\pm$0.05   \\
        & 200  & 0.47$\pm$0.02 & 0.98$\pm$0.02 & 0.11$\pm$0.05   \\
        & 300  & 0.41$\pm$0.02 & 0.97$\pm$0.01 & 0$^{+0.09}$     \\
        & 400  & 0.34$\pm$0.04 & 0.96$\pm$0.01 & 0$^{+0.08}$     \\
\hline
Cu$|$Co & 300  & 0.45$\pm$0.03 & 0.95$\pm$0.02 & 0.11$\pm$0.04   \\
\end{tabular}
\end{ruledtabular}
\end{table}

The Cu$|$Py interface parameters we calculate at various temperatures are listed in \cref{TableIV}. The low temperature (T=100K) value of $AR_{\rm I}=0.5~{\rm f}\Omega~{\rm m}^2$ that we calculate is in reasonable agreement with the low-temperature experimental value of 0.26~${\rm f}\Omega~{\rm m}^2$ \cite{Dassonneville:jap10}. The discrepancy may be attributed to unknown microscopic interface disorder in the experimental samples. Accurate measurements of the interface resistance depend on being able to separate bulk and interface contributions clearly but these are usually strongly entangled \cite{Bass:jmmm16}. The calculated value of $\gamma=0.97\pm0.01$ is also larger than the experimental value $0.7$ \cite{Dassonneville:jap10}. Improved characterization of the experimental interface structures are necessary to make progress. The SML $\delta$ is approximately 0.16$\pm$0.05 at 100 K and becomes negligible at room temperature indicating that the Cu$|$Py interface is transparent to a spin current. As the interfacial counterpart of bulk spin-flip diffusion length, the SML arises microscopically from the spin-flip scattering at the interface, which is usually induced by the enhanced spin-orbit interaction owing to the broken translational symmetry. With increasing temperature, spin fluctuation in the FM provides an additional source of spin-flip scattering. Nevertheless, the temperature-induced spin fluctuation also lowers the spin polarization parameter $\beta$ in bulk \cite{Gupta:prl20,Gupta:prb21}. Overall, the calculated SML for Cu$|$Py exhibits a monotonic decrease with increasing temperature.

\begin{figure*}[t]
\includegraphics[width=17.9cm]{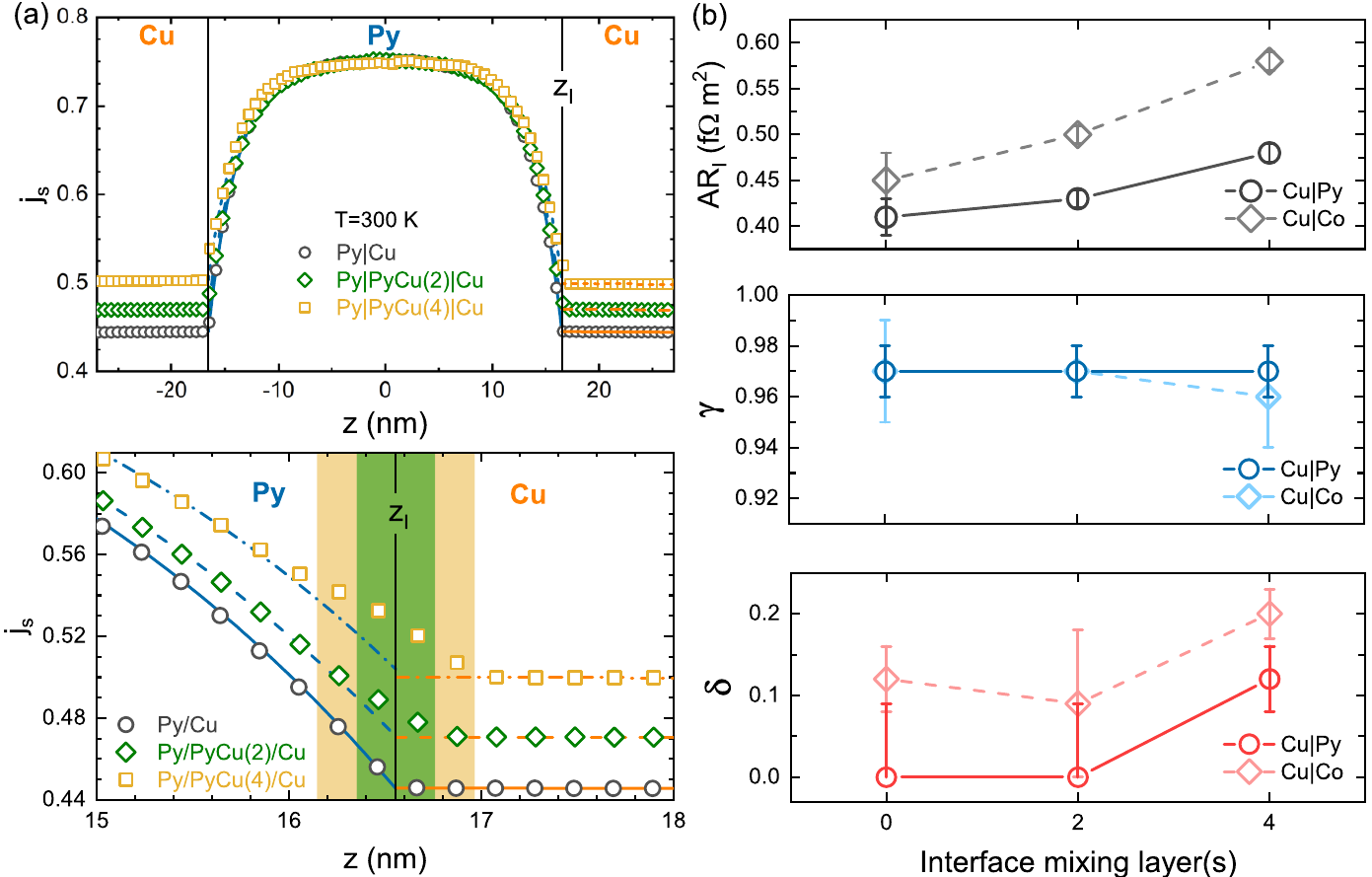}
\caption{(a) Spin currents $j_s(z)$ calculated for sharp (grey circles) and mixed Cu$|$Py interfaces with two (green diamonds) and four (yellow squares) atomic layers of mixed interface modelled as Py$_{50}$Cu$_{50}$ alloy. Thermal lattice and spin disorder are  chosen to reproduce experimental room temperature bulk resistivities. The lower panel shows an exploded view of the right Py$|$Cu interface. The solid, dashed and dash-dotted curves are fits to equations \eqref{eq:js} in Py (blue) and \eqref{eq:jsnm} in Cu (orange) for the sharp interface (vertical line at $z_{\rm I}$), two (vertical green shading) and four (vertical yellow shading) mixed interface layers, respectively. (b) Interface parameters $AR_{\rm I}$ (grey), $\gamma$ (blue) and $\delta$ (red) with $N=0,2,4$ mixed interface layers for Cu$|$Py (circles, solid lines) and Cu$|$Co (diamonds, dashed lines) interfaces.
}
\label{fig7}
\end{figure*}
	
\subsubsection{Cu$|$Co}

We repeat the above calculations at room temperature replacing Py by Co which has a higher Curie temperature and no intrinsic alloy disorder. The calculated interface parameters are included in Fig.~\ref{fig6}. A value of $AR_{\rm Cu|Co}=0.45\pm0.03~{\rm f}\Omega~{\rm m}^2$ is found which can be compared with the low-temperature experimental value of $0.20~{\rm f}\Omega~{\rm m}^2$ \cite{Dassonneville:jap10} and previously calculated value of $0.31~{\rm f}\Omega~{\rm m}^2$ without spin-orbit coupling and thermal disorder \cite{Xia:prb01}. We find a value of $\gamma=0.95\pm0.02$ that is larger than the experimental value of $\gamma=0.77\pm0.04$ \cite{Dassonneville:jap10}. The room temperature SML, $\delta=0.11\pm0.04$, for the Cu$|$Co interface compares reasonably  with the low-temperature experimental value $\delta=0.33_{-0.08}^{+0.03}$ \cite{Dassonneville:jap10}. The values of $AR_{\rm I}$ and $\delta$ we calculate for the Cu$|$Co interface are slightly larger than the corresponding values for the Cu$|$Py interface because of the greater order of Co. We found analogous results for Pt$|$Py and Pt$|$Co interfaces \cite{Gupta:prl20}. $\gamma_{\rm Cu|Co}$ is nearly the same as $\gamma_{\rm Cu|Py}$ indicating the strong spin-filtering effect of both interfaces.

\subsection{Interface mixing}
\label{ssec:intermix}

Even in the best experimental samples, the interfaces will almost certainly not be as well ordered as those we have considered so far. In the process of growing thin layers, the kinetic energy of the deposited atoms will lead to interface mixing, rendering the interfaces less sharp \cite{Bass:jmmm16}. We model the mixing of an A$|$B interface by completely mixing one (or two) layers on either side of the interface that then leads to two (or four) layers with the composition A$_{50}$B$_{50}$. The lattice and spin disorder is assumed to be unchanged, as are the atomic sphere potentials. The spin currents that result from these calculations for Cu$|$Py$|$Cu structures with intermixed interfaces are shown in \cref{fig7}(a). They are fitted in Py and Cu using equations \eqref{eq:js} and \eqref{eq:jsnm}, respectively and extrapolated to the Py$|$Cu interface at $z_{\rm I}$ to determine $j_{s}(z_{\rm I}-\eta)$ on the Py side and $j_{s}(z_{\rm I}+\eta)$ on the Cu side, \cref{fig7}(a), lower panel. The interface resistance $AR_{\rm I}$ is determined in separate calculations and finally $\delta$ and $\gamma$ are extracted by solving equations \eqref{eq:nmfm}. The results are shown for both Cu$|$Py and Cu$|$Co interfaces in \cref{fig7}(b). $AR_{\rm I}$ is seen to increase monotonically with increasing thickness of the mixed interface layer because of the strong scattering by alloy disorder. The interface spin asymmetry parameter $\gamma$ is not changed by the intermixing. The majority-spin potentials of Co, and of both Ni and Fe in Py, are perfectly matched to the potential of Cu while the minority-spin potentials all differ. For this reason, both Cu$|$Py and Cu$|$Co interfaces exhibit strong spin filtering and this effect is not significantly weakened by mixing the magnetic and Cu atoms. 

The effect of interface mixing on the SML for Cu$|$Py and Cu$|$Co interfaces is complex. First, the stronger interface scattering by the interface alloy enhances the spin flipping, as we found for the nonmagnetic Au$|$Pt interface \cite{Gupta:prl20}. For a Cu$|$FM interface, alloying reduces the magnetic order on the FM side and this reduces the SML by analogy with the reduction we found for Pt$|$Co and Pt$|$Py interfaces on increasing the temperature \cite{Gupta:prl20}. Competition between the two effects results in the (slightly) nonmonotonic dependence of $\delta$ on the thickness of the interface alloy layer at Cu$|$Py and Cu$|$Co interfaces shown in \cref{fig7}(b). This nonmonotonic behavior is clearer for Cu$|$Co since Co has a higher degree of magnetic order than Py.

\subsection{Inserting copper at a FM$|$NM interface}
\label{ssec:Cu_insert}

Cu is commonly used in experiments as a spacer material between a heavy metal like Pt with a large spin susceptibility and a FM metal, to prevent magnetism being induced in Pt by the proximity effect. Because of its weak spin-orbit interaction and correspondingly long SDL, a thin layer of Cu is usually assumed to have no effect on a spin current thus making the interpretation of experiments simpler. However, though bulk Cu may have little effect on a spin current, insertion of a Cu layer between a FM metal and Pt replaces the singel FM$|$Pt interface with two different interfaces, namely, FM$|$Cu and Cu$|$Pt interfaces whose effect on a spin current is, at best, poorly known. Here we consider Py and Co as typical FM metals and Pt as a typical heavy metal to investigate the influence of inserting Cu layers on the spin memory loss.
	
\begin{figure}[!t]
\includegraphics[width=8.6cm]{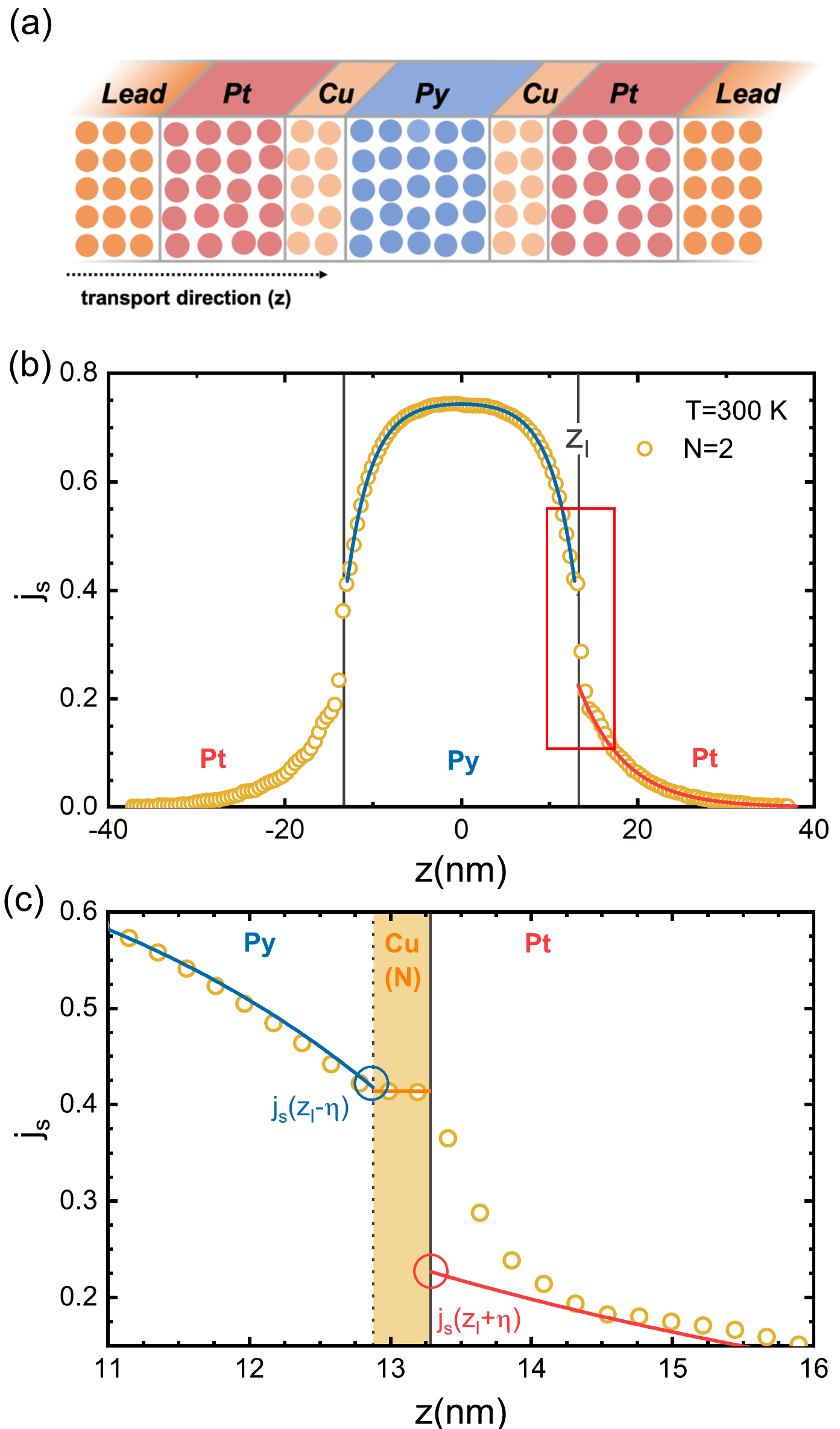}
\caption{(a) Sketch of a Pt$|$Cu$|$Py$|$Cu$|$Pt multilayer sandwiched between semi-infinite ballistic Cu leads. All atoms in the scattering region are displaced from their equilibrium positions on fcc lattices to simulate room temperature lattice disorder. Spin disorder in Py is included in the same way as before. 
(b) Spin current $j_s(z)$ calculated for a Pt$|$Cu(2)$|$Py$|$Cu(2)$|$Pt multilayer where two atomic layers of Cu are inserted at the Py$|$Pt interfaces. An exploded view of the discontinuity inside the red rectangle about the right-hand Py$|$Cu$|$Pt interface is shown in (c) where the solid lines represent the piecewise fitting in Py and Pt using the solution to the spin diffusion equations \eqref{eq:js} and ~\eqref{eq:jsnm} respectively. These fits are extrapolated to a position $z_{\rm I}$ between the Py$|$Cu and Cu$|$Pt interface to obtain the values $j_{s}(z_{\rm I}-\eta)$ and $j_{s}(z_{\rm I}+\eta)$, respectively, indicated schematically by the large open circles. Shifting the position $z_{\rm I}$ about inside Cu has a negligible effect on the values of $\gamma$ and $\delta$ we extract because of the thinness of Cu. The Cu insert is considered as a ``compound'' Py$|$Pt interface.
}
\label{fig8}
\end{figure}	

As shown schematically in \cref{fig8}(a), we calculate the spin current distribution at 300~K for a Pt$|$Cu$|$Py$|$Cu$|$Pt multilayer when a charge current is passed through it. The thickness of the Py (Pt) layers is 26~nm (25~nm) and we consider $N=0$, 1 and 2 atomic layers of Cu. The spin current in the transport direction ($z$) that we calculate for a symmetric Pt$|$Cu($N$)$|$Py$|$Cu($N$)$|$Pt multilayer is shown in \cref{fig8}(b) for $N=2$. Both the saturated plateau in the center of Py and the vanishing spin current at the interfaces between Pt and the leads confirm that Py and Pt are sufficiently thick and satisfy the boundary condition $j_s(\pm\infty)=0$. 
We fit $j_s(z)$ using the spin diffusion equations in Py and Pt and use these fits to extrapolate $j_s(z)$ on the Py side to the Py$|$Cu interface to calculate $j_{s}(z_{\rm I}-\eta)$ and on the Pt side to the Cu$|$Pt interfaces to calculate $j_{s}(z_{\rm I}+\eta)$, respectively as illustrated in the exploded plot in \cref{fig8}(c). The somewhat arbitrary choice of $z_{\rm I}$ within the Cu insert has a negligible effect on the values of $\delta$ and $\gamma$ that we extract because of the thinness of Cu. The SML determined in this way accounts for the spin-flipping at the ``compound interface'' between Py and Pt.

\begin{figure}[t]
\includegraphics[width=\columnwidth]{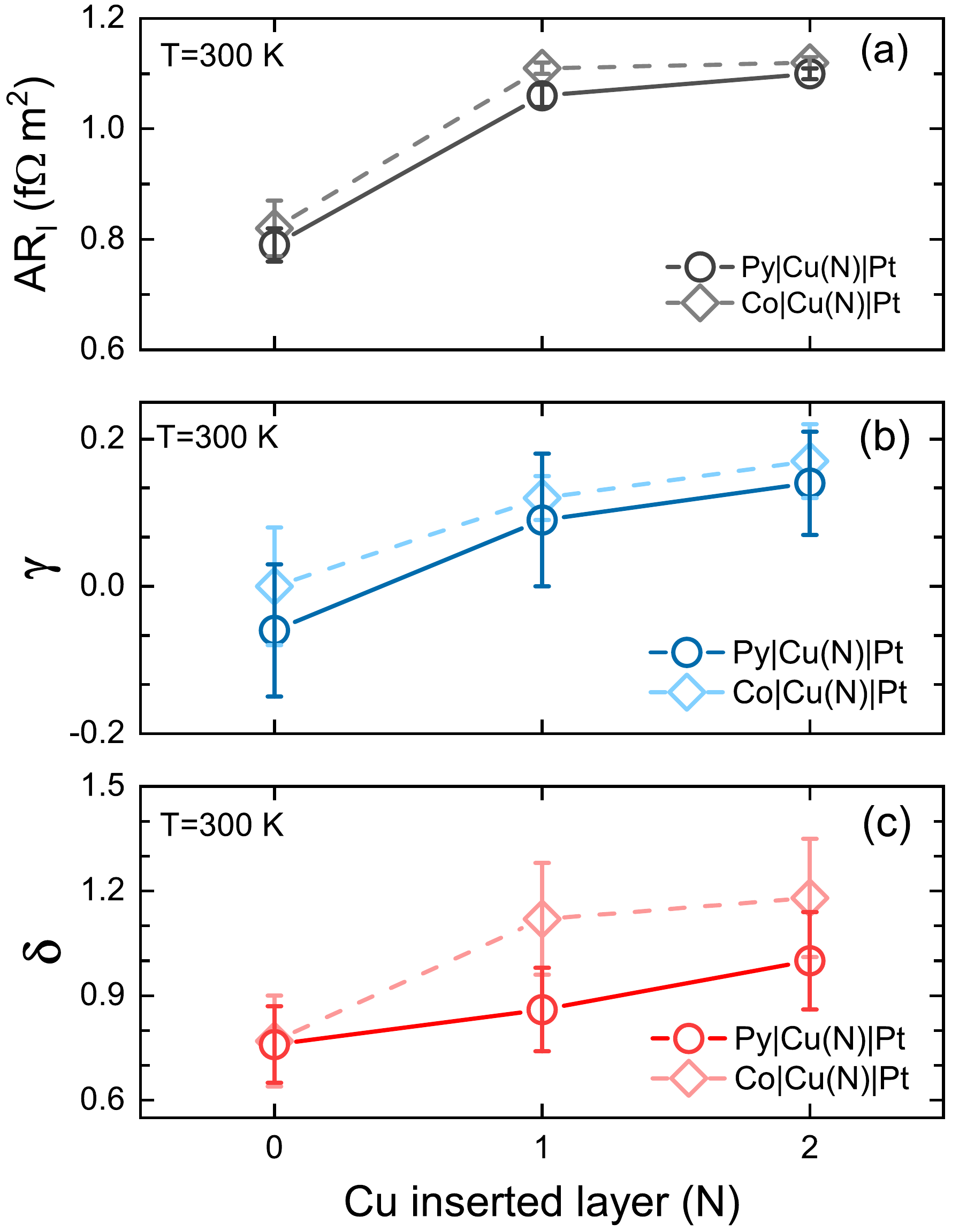}
\caption{Calculated interface resistance $AR_{\rm I}$ (a), spin-asymmetry parameter $\gamma$ (b), and SML $\delta$ (c) when $N$ atomic layers of Cu ($N=$0,1,2) are inserted into the Py$|$Pt (circles) and Co$|$Pt (diamonds) interfaces at room temperature. The solid and dashed lines are guides to the eye.}
\label{fig9}
\end{figure}

Using the above scheme, we calculate the interface parameters $AR_{\rm I}$, $\gamma$ and $\delta$ for Py$|$Cu$|$Pt and Co$|$Cu$|$Pt interfaces at room temperature and summarize the results in \cref{fig9} and Table~\ref{table5}. The Cu insert increases $AR_{\rm Py|Pt}$ to between 1.06 and $1.11~{\rm f}\Omega\,{\rm m}^2$, which is very close to the sum of the two interface resistances connected in series, $AR_{\rm Py|Cu}+AR_{\rm Cu|Pt}=1.05\pm0.04~{\rm f}\Omega~{\rm m}^2$. Because Cu is so thin and conductive, its contribution to $AR_{\rm I}$ can be neglected. The parameter $\gamma$ is small for both the Py$|$Pt and Co$|$Pt interfaces but increases with increasing Cu thickness. This is because the Py$|$Cu and Co$|$Cu interfaces have strong spin filtering effects. The scattering rate for minority-spin electrons is enhanced by the Cu insert leading to an increase in $\gamma$. The dependence of $AR_{\rm I}$ and $\gamma$ on the thickness of Cu is the same for both Py$|$Cu$|$Pt and Co$|$Cu$|$Pt interfaces.

\begin{table}[t]
\caption{Calculated interface parameters $AR_{\rm I}$, $\gamma$ and $\delta$ for $N$ atomic layers of Cu ($N=$0,1,2) inserted into the Py$|$Pt and Co$|$Pt interfaces at room temperature.}
\begin{ruledtabular}
\begin{tabular}{lcccc}
                  & $N$ & $AR_{\rm I}(\rm f\Omega\,{\rm m^2})$ 
                                        & $\gamma$       & $\delta$      \\
\hline
Py$|$Cu($N$)$|$Pt &  0  & 0.79$\pm$0.03 & -0.06$\pm$0.09 & 0.76$\pm$0.11 \\
                  &  1  & 1.06$\pm$0.02 &  0.09$\pm$0.09 & 0.86$\pm$0.12 \\
                  &  2  & 1.11$\pm$0.01 &  0.14$\pm$0.07 & 1.00$\pm$0.14 \\
\hline
Co$|$Cu($N$)$|$Pt &  0  & 0.82$\pm$0.05 &  0.00$\pm$0.08 & 0.77$\pm$0.13 \\
                  &  1  & 1.11$\pm$0.01 &  0.12$\pm$0.03 & 1.12$\pm$0.16 \\
                  &  2  & 1.12$\pm$0.01 &  0.17$\pm$0.05 & 1.18$\pm$0.17 \\
\end{tabular}
\end{ruledtabular}
\label{table5}
\end{table}

We finally consider the SML with and without the Cu insert. Without it, $\delta_{\rm Py|Pt}=0.76\pm0.11$ at room temperature. As shown in \cref{table5} and Fig.~\ref{fig9}, inserting Cu increases $\delta_{\rm Py|Pt}$ slightly to $0.86\pm0.12$ for $N=1$ and to $1.00\pm0.14$ for $N=2$ compared to $\delta_{\rm Py|Cu}+\delta_{\rm Cu|Pt}=(0.00^{+0.09})+(0.77\pm0.04)=0.77\pm0.10$ for the two separate interfaces.
For the Co$|$Pt interface, the SML increases from $\delta_{\rm Co|Pt}=0.77\pm0.13$ to $1.12\pm0.16$ for $N=1$ and $1.18\pm0.17$ for $N=2$. The sum of the room temperature SML values for the individual interfaces, $\delta_{\rm Co|Cu}+\delta_{\rm Cu|Pt}=(0.11\pm0.04)+(0.77\pm0.04)=0.88\pm0.06$, is substantially lower than the SML we find for the compound Co$|$Cu$|$Pt interface. 
Thus, contrary to the expectation that separating the FM and heavy metal should enhance the interface transparency for a spin current \cite{Zhu:prl21}, we find that it increases the SML.

\section{Conclusion}
\label{sec:conclusion}

We have calculated the semiclassical spin transport parameters for Cu$|$Pt, Cu$|$Pd, Cu$|$Py and Cu$|$Co interfaces at finite temperature within the adiabatic approximation using first-principles scattering theory \cite{Starikov:prb18} and a recently developed planar-averaged local current scheme \cite{Wesselink:prb19}. The dependence of the interface parameters on temperature and interface atomic mixing is studied systematically. The SML at a Cu$|$Pt interface is comparable to what we found for an incommensurate Au$|$Pt interface in \cite{Gupta:prl20} which we attribute to the similarity of the free-electron-like electronic structures of the noble metals Cu and Au. For Cu$|$Py and Cu$|$Co interfaces, both the interface resistance and SML are found to decrease monotonically with temperature. The SML becomes negligibly small at room temperature. By analogy with the Pt$|$FM interfaces, the SML is larger for Cu$|$Co than for Cu$|$Py because unlike Py,  Co does not have alloy disorder and has a higher Curie temperature. Inserting a thin layer of Cu in the Py$|$Pt or Co$|$Pt interfaces increases the SML.  Since Cu is widely used as a transport channel in nonlocal spin valves, our calculated values of interface transport parameters for the Cu$|$NM and Cu$|$FM interfaces should be  very useful reference data for experimental studies. 

Where comparison can be made, the results we calculate are in reasonable agreement with published experimental values. The sophistication of our calculations is such that where discrepancies exist, the first issue that much be examined is the validity of the structural models we use. Our computer codes make it possible to examine many types of interface disorder but at present more information is required from experiment to motivate more extensive theoretical investigations than the present one. Where the experimental data were obtained from low-temperature models, the onus is on our experimental colleagues to provide us with information about the disorder that leads to diffusive behaviour at low temperatures. 

\acknowledgements
This work was financially supported by  the National Natural Science Foundation of China (Grants No. 12174028 and No. 11734004), the Recruitment Program of Global Youth Experts and by the ``Nederlandse Organisatie voor Wetenschappelijk Onderzoek'' (NWO) through the research programme of the former ``Stichting voor Fundamenteel Onderzoek der Materie,'' (NWO-I, formerly FOM). K.G. acknowledges funding from the Shell-NWO/FOM “Computational Sciences for Energy Research” PhD program (CSER-PhD; nr.~i32; project number 13CSER059). The work was also supported by the Royal Netherlands Academy of Arts and Sciences (KNAW).

\appendix
\section{The Valet-Fert formalism for a NM$|$FM bilayer}
\label{sec:app}

The RT value of $l_{\rm Pt} \sim 5.3 \,$nm makes it possible to realize the boundary condition $j_s(\pm \infty) \rightarrow 0$ for Pt$|$FM$|$Pt scattering regions that can be handled in practical calculations \cite{Gupta:prl20, Gupta:prb21}. The large value of $l_{\rm Cu} \sim 502 \,$nm makes this impossible for Cu$|$FM$|$Cu raising doubts about the value of $\delta$ calculated in \cref{ssec:CuFM}. In this Appendix we derive semiclassical transport equations based on the Valet-Fert formalism for a NM$|$FM bilayer with a spin-polarized current incident from the left lead for which the boundary condition is $j_s(-\infty)=\pm1$. Using this alternative boundary condition to determine the SML for Cu$|$FM interfaces yields the same numerical values of $\delta$ as the $j_s(\pm \infty) \rightarrow 0$ boundary condition. We demonstrate that the two different calculation schemes yield the same results within the numerical accuracy for a Pt$|$Py interface. 

We follow the standard procedure used in the literature \cite{Baxter:jap99, Eid:prb02} and treat the interface (I) as an artificial bulklike material with resistivity $\rho_{\rm I}$, SDL $l_{\rm I}$ and finite thickness $t$ that are related to an interface resistance and SML as $AR_{\rm I}=\rho_{\rm I}t$ and $\delta=t/l_{\rm I}$. The general solutions \eqref{eq:solution} and \eqref{eq:js} to the spin diffusion equations and Ohm's law  have the same forms in the NM, I (interface), and FM regions 
\begin{eqnarray}
\mu_{si}(z)&=&A_ie^{z/l_i}+B_ie^{-z/l_i},\label{eqa:mus} \\
j_{si}(z)  &=&\beta_i-\frac{1-\beta_i^2}{2ej\rho_i l_i}\left(A_i e^{z/l_i}-B_i e^{-z/l_i}\right).
\label{eqa:js}
\end{eqnarray}

If a fully polarized spin current is injected from an artificial half-metallic left lead into a diffusive NM$|$FM bilayer, one has the boundary condition $j_{s,\rm NM}(0)=\pm1$, where the sign $+$($-$) indicates that the polarization is parallel (antiparallel) to the current polarization direction in the FM metal. Substituting this boundary condition at $z=0$ into \eqref{eqa:js} and considering $\beta=0$ in the NM metal, we find
\begin{equation}
A_{\rm NM}=B_{\rm NM}\mp2ej\rho_{\rm NM}l_{\rm NM}.
\end{equation}
The above equation can be substituted into \eqref{eqa:mus} and \eqref{eqa:js} to eliminate the coefficients $A_{\rm NM}$ and $B_{\rm NM}$. Eventually we arrive at the following relation between the spin accumulation $\mu_{s,\rm NM}$ and the normalized spin current $j_{s,\rm NM}$ in the NM metal
\begin{equation}
\mu_{s,\rm NM}=\frac{2ej\rho_{\rm NM}l_{\rm NM}}{\sinh(z/l_{\rm NM})}
                 \big[\pm 1-j_{s, \rm NM}(z)\cosh(z/l_{\rm NM})\big].
\label{eqa:munm}
\end{equation}
In the FM metal, the boundary condition $j_{s,\rm FM}(+\infty)=\beta$ results in $A_{\rm FM}=0$. The other coefficient $B_{\rm FM}$ can be eliminated using \eqref{eqa:mus} and \eqref{eqa:js} and the spin accumulation in the FM metal reads
\begin{equation}
\mu_{s,\rm FM}(z)=\frac{2ej\rho_{\rm FM} l_{\rm FM}}{1-\beta^2}
\big[j_{s,{\rm FM}}(z)-\beta \big]. 
\label{eqa:mufm}
\end{equation}

Since the interface is replaced by an artificial bulklike material with a finite thickness $t$, the spin accumulation $\mu_{s}(z)$ and spin current $j_{s}(z)$ are continuous everywhere. At the NM$|$I boundary $z=z_{\rm I}$ and at the I$|$FM boundary $z=z_{\rm I}+t$, the spin accumulation $\mu_s$ and spin current $j_s$ are both continuous,
\begin{widetext}
\begin{eqnarray}
\mu_{s,\rm NM}(z_{\rm I})&=&\mu_{s,\rm I}(z_{\rm I})=A_{\rm I}e^{z_{\rm I}/l_{\rm I}}+B_{\rm I}e^{-z_{\rm I}/l_{\rm I}},\label{eqa:mu1}\\
\mu_{s,\rm FM}(z_{\rm I}+t)&=&\mu_{s,\rm I}(z_{\rm I}+t)=A_{\rm I}e^{(z_{\rm I}+t)/l_{\rm I}}+B_{\rm I}e^{-(z_{\rm I}+t)/l_{\rm I}},\label{eqa:mu2}\\
j_{s,\rm NM}(z_{\rm I})&=&j_{s,\rm I}(z_{\rm I})=\gamma-\frac{1-\gamma^2}{2ej\rho_I l_{\rm I}}\left(A_{\rm I}e^{z_{\rm I}/l_{\rm I}}-B_{\rm I}e^{-z_{\rm I}/l_{\rm I}}\right),\label{eqa:js1}\\
j_{s,\rm FM}(z_{\rm I}+t)&=&j_{s, \rm I}(z_{\rm I}+t)=\gamma-\frac{1-\gamma^2}{2ej\rho_I l_{\rm I}}\left(A_{\rm I}e^{(z_{\rm I}+t)/l_{\rm I}}-B_{\rm I}e^{-(z_{\rm I}+t)/l_{\rm I}}\right).\label{eqa:js2}
\end{eqnarray}
\eqref{eqa:mu1} and \eqref{eqa:mu2} can be used to express the coefficients $A_{\rm I}$ and  $B_{\rm I}$ as functions of the spin accumulation at the NM$|$I and I$|$FM boundaries as
\begin{eqnarray}
A_{\rm I}&=&\frac{\mu_{s,\rm NM}(z_{\rm I})e^{-t/l_{\rm I}}-\mu_{s,\rm FM}(z_{\rm I}+t)}{e^{z_{\rm I}/l_{\rm I}}\big(e^{-t/l_{\rm I}}-e^{t/l_{\rm I}}\big)}, \label{eq:ai}\\
B_{\rm I}&=&\frac{\mu_{s,\rm NM}(z_{\rm I})e^{t/l_{\rm I}}-\mu_{s,\rm FM}(z_{\rm I}+t)}{e^{-z_{\rm I}/l_{\rm I}}\big(e^{t/l_{\rm I}}-e^{-t/l_{\rm I}}\big)}. \label{eq:bi}
\end{eqnarray}
Substituting \eqref{eq:ai} and \eqref{eq:bi} into \eqref{eqa:js1} and \eqref{eqa:js2}, we find two equations containing $\mu_s$ and $j_s$ at the boundaries $z=z_{\rm I}$ and $z=z_{\rm I}+t$. The spin accumulation $\mu_s$ is eliminated using \eqref{eqa:munm} and \eqref{eqa:mufm} and finally, we take the limit $t\rightarrow0$ to arrive at  equations \eqref{eq:nmfmv2} that only depend on spin currents
\begin{subequations}
\small
\label{eqa:delta12}
\begin{eqnarray}
j_{s,\rm NM}(z_{\rm I})&=&\gamma-\frac{(1-\gamma^2)\delta}{AR_{\rm I}\sinh\delta}
\left\{\frac{\rho_{\rm FM}l_{\rm FM}}{1-\beta^2}\left[j_{s,\rm FM}(z_{\rm I})-\beta\right]-\rho_{\rm NM}l_{\rm NM}\left[\pm {\rm csch}\left(\frac{z_{\rm I}}{l_{\rm NM}}\right)-j_{s,\rm NM}(z_{\rm I}){\rm coth}\left(\frac{z_{\rm I}}{l_{\rm NM}}\right)\right]\cosh\delta\right\}, \;\;\;\;\; \label{eqa:delta12a}\\
j_{s,\rm FM}(z_{\rm I})&=&\gamma-\frac{(1-\gamma^2)\delta}{AR_{\rm I}\sinh\delta}
\left\{\frac{\rho_{\rm FM}l_{\rm FM}}{1-\beta^2}\left[j_{s,\rm FM}(z_{\rm I})-\beta\right]\cosh\delta-\rho_{\rm NM}l_{\rm NM}\left[\pm {\rm csch}\left(\frac{z_{\rm I}}{l_{\rm NM}}\right)-j_{s,\rm NM}(z_{\rm I}){\rm coth}\left(\frac{z_{\rm I}}{l_{\rm NM}}\right)\right]\right\}. \label{eqa:delta12b}
\end{eqnarray}
\end{subequations}
Here $j_{s,\rm NM}(z_{\rm I}) \equiv j_s(z_{\rm I}-\eta)$ in \eqref{eq:nmfmv2} and similarly $j_{s,\rm FM}(z_{\rm I}) \equiv j_s(z_{\rm I}+\eta)$ and we have already made use of the relations $\rho_{\rm I}t=AR_{\rm I}$ and $l_{\rm I}/t=\delta$.  

For the special case $\beta=\gamma=0$, the NM$|$FM interface becomes an NM$|$NM$'$ interface and \eqref{eqa:delta12a} and \eqref{eqa:delta12b} reduce to
\begin{subequations}
\small
\label{eqa:delta13}
\begin{eqnarray}
j_{s,\rm NM}(z_{\rm I})&=&\frac{\delta}{AR_{\rm I}\sinh\delta}\left\{-\rho_{\rm NM'}l_{\rm NM'}j_{s,\rm NM'}(z_{\rm I})+\rho_{\rm NM}l_{\rm NM}\left[\pm {\rm csch}\left(\frac{z_{\rm I}}{l_{\rm NM}}\right)-j_{s,\rm NM}(z_{\rm I}){\rm coth}\left(\frac{z_{\rm I}}{l_{\rm NM}}\right)\right]\cosh\delta\right\},  \label{eqa:delta13a}  \\
j_{s,\rm NM'}(z_{\rm I})&=&\frac{\delta}{AR_{\rm I}\sinh\delta}\left\{-\rho_{\rm NM'}l_{\rm NM'}j_{s,\rm NM'}(z_{\rm I})\cosh\delta+\rho_{\rm NM}l_{\rm NM}\left[\pm {\rm csch}\left(\frac{z_{\rm I}}{l_{\rm NM}}\right)-j_{s,\rm NM}(z_{\rm I}){\rm coth}\left(\frac{z_{\rm I}}{l_{\rm NM}}\right)\right]\right\}. \label{eqa:delta13b}
\end{eqnarray}
\end{subequations}
Eliminating $\rho_{\rm NM}l_{\rm NM}$ from the above two equations, we obtain 
\begin{equation}
\frac{j_{s,\rm NM}(z_{\rm I})}{j_{s,\rm NM'}(z_{\rm I})}=\cosh\delta+\frac{\rho_{\rm NM'}l_{\rm NM'}}{AR_{\rm I}}\delta \sinh\delta,
\end{equation}
and reproduce \eqref{eq:nmnm} in Sec.~\ref{ssec:nmnmp}. 
\end{widetext}

\subsubsection*{Pt$|$Py bilayer}

\begin{figure}[t]
\includegraphics[width=\columnwidth]{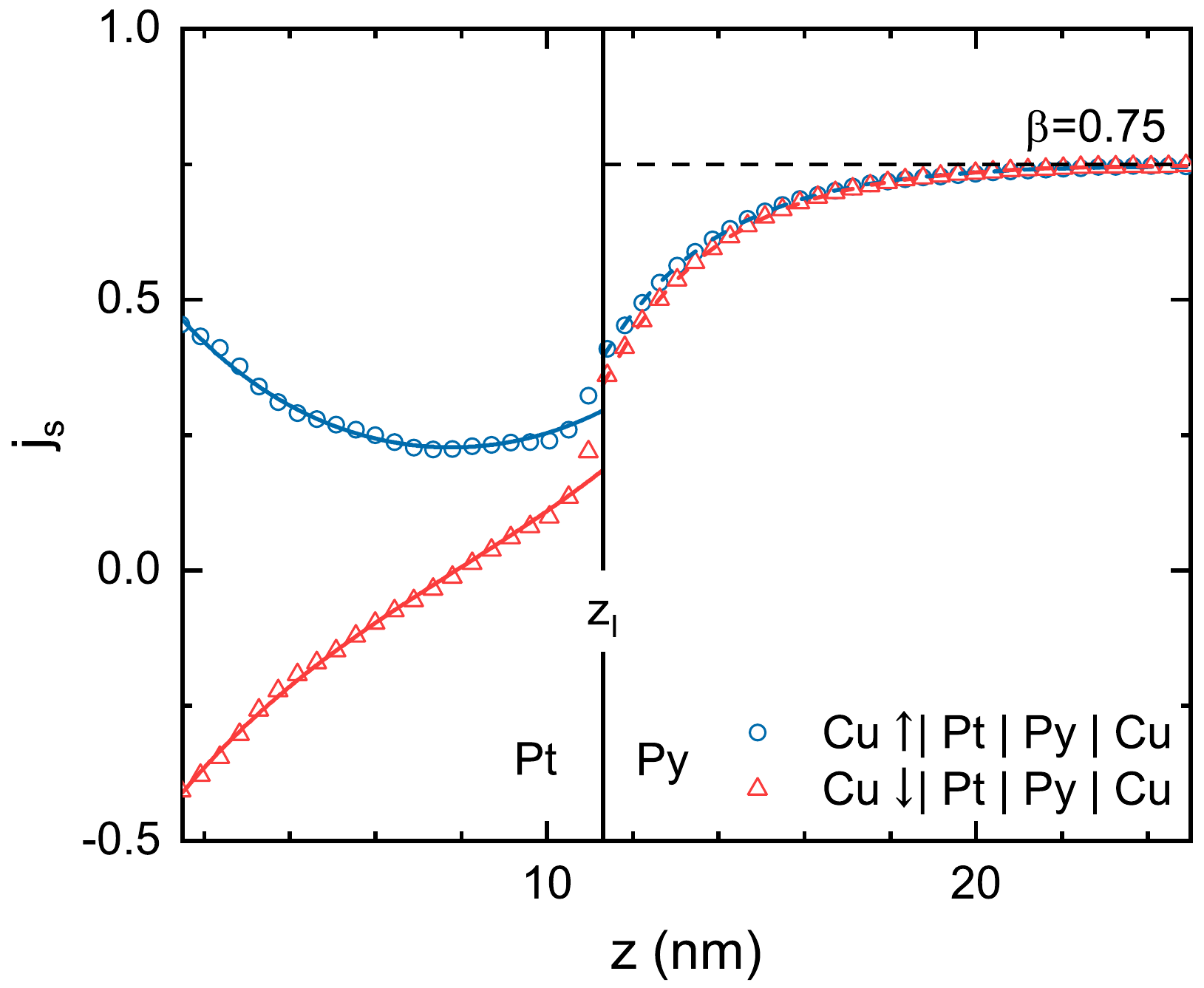}
\caption{Spin currents $j_{s}(z)$ calculated inside a diffusive Pt(10 nm)$|$Py(15 nm) bilayer at room temperature. Fully polarized spin-up (blue circles) and spin-down (red triangles) currents injected from artificial half-metallic Cu($\uparrow$) and Cu($\downarrow$) left leads, respectively. The magnetization of Py is set as its polarization $\beta>0$ indicated by the horizontal dashed line. Each data is an average over 20 random configurations of thermal lattice and spin disorder. The solid and dashed lines are fits using (\ref{eqa:js}) in Pt and Py, respectively, with appropriate boundary conditions. 
}
\label{fig10}
\end{figure}

To examine the validity of extracting the SML $\delta$ and interface polarization $\gamma$ for a NM$|$FM interface with a fully polarized current injected from the NM side, we take Pt$|$Py as an example and compare the numerical results we find with those obtained by passing an unpolarized current through a NM$|$FM$|$NM trilayer structure \cite{Gupta:prl20, Gupta:prb21}. We construct a diffusive Pt(10 nm)$|$Py(15 nm) bilayer at room temperature with thermal lattice and spin disorder. The lattice mismatch between the two fcc metals is accommodated using a $2\sqrt{13}\times 2\sqrt{13}$ supercell of (111) oriented Pt matched to a $8\times8$ supercell of (111) Py. The bilayer is then sandwiched between Cu leads whose lattice constant is chosen to be the same as that of Py. The minority-spin (or majority-spin) potential of Cu in the left lead is artificially increased by 1 Rydberg so that all the incoming Bloch states have pure spin character and the charge current injected from the left Cu lead is fully spin-polarized. In the transport calculation, the 2D Brillouin zone of the matched lateral supercell is sampled using a $28\times28$ $k$-mesh corresponding to a $224\times224$ sampling of a unit cell of fcc Py. The results we show are obtained by averaging over twenty random configurations of thermally disordered Pt$|$Py bilayer. Note that the Py is sufficiently thick that the right-hand lead does not affect $j_s$ at the Pt$|$Py interface plotted in Fig.~\ref{fig10}.

The plane-averaged spin current $j_s(z)$ we obtain for the Pt$|$Py bilayer is shown in \cref{fig10}. Using \eqref{eqa:js}, it can be fitted piecewise in Pt (solid lines) and in Py (dashed lines) and the values required at the interface are obtained by extrapolating these fits. On injecting a spin current with positive polarization into Pt, $j_s(-\infty)=1$ (blue symbols and lines), we find $j_{s}(z_{\rm I}-\eta)=0.28\pm0.01$ and $j_{s}(z_{\rm I}+\eta)=0.39\pm0.01$ by extrapolation. Substituting these values into \eqref{eqa:delta12a} and \eqref{eqa:delta12b} together with the independently determined bulk parameters, $\rho_{\rm Pt}=10.7\pm0.03\,\mu\Omega\,{\rm cm}$, $l_{\rm Pt}=5.25\pm0.03$~nm, $\rho_{\rm Py}=15.6\pm0.02\,\mu\Omega\,{\rm cm}$, $l_{\rm Py}=2.85\pm0.02$~nm and $AR_{\rm Pt|Py}=0.79\pm0.03\,{\rm f}\Omega\,{\rm m}^2$, we finally obtain $\delta=0.65\pm0.14$ and $\gamma=-0.03\pm0.06$. 
Injecting a spin current with negative polarization, $j_s(-\infty)=-1$ (red symbols and lines), we find  $j_{s}(z_{\rm I}-\eta)=0.17\pm0.01$ and $j_{s}(z_{\rm I}+\eta)=0.34\pm0.01$ yielding $\delta=0.63\pm0.08$ and $\gamma=-0.03\pm0.07$. These values are consistent with the values calculated using the Pt$|$Py$|$Pt trilayer structure $\delta=0.76\pm0.11$ and $\gamma=-0.06\pm0.09$ within the error bars of the calculations \cite{Gupta:prl20, Gupta:prb21}. 

\end{document}